\documentclass[prd,tightenlines,nofootinbib,eqsecnum,superscriptaddress,twocolumn]{revtex4-2}

\usepackage{amsmath}
\usepackage{amsfonts}
\usepackage{amssymb}
\usepackage{amsthm}
\usepackage{amstext}
\usepackage{graphicx}
\usepackage{xcolor}
\usepackage{comment}
\usepackage{multirow}

\usepackage{mathrsfs}
\usepackage{bm}
\usepackage{array}
\usepackage{booktabs}
\usepackage{verbatim}
\usepackage{pagecolor}
\usepackage{enumerate}
\usepackage{mathtools}
\usepackage{orcidlink}
\usepackage[normalem]{ulem}
\usepackage{scalerel}
\usepackage{textcomp}
\usepackage{url}
\usepackage{enumitem}

\usepackage{hyperref}
\hypersetup{
    colorlinks=true,
    linktocpage,
    unicode,
    linkcolor=blue,
    filecolor=magenta,
    urlcolor=cyan
}

\usepackage{afterpage}

\newcolumntype{C}{>{$}c<{$}}
\AtBeginDocument{
\heavyrulewidth=.08em
\lightrulewidth=.05em
\cmidrulewidth=.03em
\belowrulesep=.65ex
\belowbottomsep=0pt
\aboverulesep=.4ex
\abovetopsep=0pt
\cmidrulesep=\doublerulesep
\cmidrulekern=.5em
\defaultaddspace=.5em
}

\definecolor{red }{rgb}{1,0,0}
\definecolor{blue}{rgb}{0,0,1}
\definecolor{green}{rgb}{0,1,0}
\definecolor{CiteColor}{rgb}{0,0,0.35}
\definecolor{RefColor}{rgb}{0.55,0,0}
\definecolor{URLColor}{rgb}{0,0,0.35}

\DeclareSymbolFontAlphabet{\mathrsfs}{rsfs}

\hypersetup{colorlinks,citecolor=CiteColor,urlcolor=URLColor}
\hypersetup{citecolor=CiteColor}
\hypersetup{linkcolor=RefColor}

\newcommand{\beq}{\begin{equation}}
\newcommand{\eeq}{\end{equation}}
\newcommand{\quand}{\quad \text{and} \quad}

\newcommand{\ud}{\mathrm{d}}
\newcommand{\ui}{\mathrm{i}}

\newcommand{\RR}{\mathbb{R}}
\newcommand{\CC}{\mathbb{C}}

\newcommand{\EE}{\mathbb{E}}
\newcommand{\BB}{\mathbb{B}}

\newcommand{\mcM}{\mathcal{M}}

\newcommand{\mcB}{\mathcal{B}}

\newcommand{\mcE}{\mathcal{E}}

\newcommand{\muT}{\mu_{\text{T}}}
\newcommand{\cE}{c_\text{E}}
\newcommand{\cB}{c_\text{B}}

\newcommand{\ph}{\phantom{a}}
\newcommand{\hp}{\hat{p}}

\newcommand{\Hgeo}{H_0}
\newcommand{\Htidal}{H_1}
\newcommand{\Qgeo}{Q_0}
\newcommand{\Qtidal}{Q_1}
\newcommand{\lambdaH}{\lambda_\text{H}}

\begin{document}

\title{Quadrupolar tidal effects destroy the integrability of black hole geodesics:\\analytic proof and numerical evidence of chaos}

\author{Paul Ramond,\orcidlink{0000-0001-7123-0039}}\,\email{ramond@lpccaen.in2p3.fr}
\affiliation{Universit\'e de Caen Normandie, ENSICAEN, CNRS/IN2P3,\\LPC Caen UMR6534, F-14000 Caen, France}

\date{\today}

\begin{abstract}

In general relativity, the motion of a test mass around a rotating black hole is described by Kerr geodesics. Owing to the symmetries of the Kerr spacetime, these geodesics possess four constants of motion, rendering the associated Hamiltonian system integrable. This integrability underlies much of the analytical framework used to model asymmetric-mass-ratio inspirals, key sources for future gravitational-wave detectors. Real compact bodies, however, are not test masses: their internal structure couples to the background curvature.
In this work, we show that a non-spinning body endowed with a tidally induced quadrupole admits no deformation of the geodesic Carter constant that remains conserved, for generic tidal couplings and generic Kerr spin. Consequently, the leading-order tidal dynamics is generically non-integrable. 
The proof is analytic and relies on two key ingredients: a covariant Hamiltonian formulation of tidal dynamics on the same phase space as the geodesic problem, valid in arbitrary background spacetimes, and a novel relation between curvature tidal scalars and the geodesic Carter constant derived from the algebraic and Killing symmetries of Kerr spacetime. 
We complement this result with numerical diagnostics of the tidally perturbed dynamics, including Poincaré sections, Lyapunov exponents, and escape-time maps. These reveal chaotic structures in phase space, such as stochastic layers, sensitivity to initial conditions, and fractal basin boundaries, consistently with the analytic non-integrability result.

\end{abstract}

\maketitle
\tableofcontents

%%%%%%%%%%%%%%%%%%%%%%%%%%%%%%%%%%%%%%%%%%
\section{Introduction and Summary}
\label{sec:intro}
%%%%%%%%%%%%%%%%%%%%%%%%%%%%%%%%%%%%%%%%%%%

\subsection{Context}

General relativity predicts that, to a very good approximation, the trajectory of a test mass around a generic black hole is a \emph{geodesic} of the Kerr spacetime. As such, right after the discovery of the Kerr metric itself \cite{Kerr:1963ud}, the 1960s saw a fast development in the study of Kerr geodesics, marked with Carter's discovery of the eponymous constant of motion \cite{Carter.68}. Carter's finding was soon shown to be tied to a fundamental, hidden symmetry of the Kerr spacetime, encoded in a Killing--St\"ackel tensor and, more fundamentally, in the Killing--Yano tensor from which it is built \cite{WalkPen.70,Floyd.73,HughSomm.73}. These pioneering results unlocked a wide variety of developments right at the advent of the golden age of general relativity. Many studies were conducted to solve the geodesic equations in various orbital configurations and to classify them \cite{LePe.08,CoLiLo.22,StWa.20}, both for null geodesics \cite{Gralla:2019ceu,Cieslik:2023qdc,Cieslik:2022uki} and for timelike ones \cite{Mino:2003yg,DraHug.04,Schm.02,FuHi.09,VdM.20}, which are useful respectively for black-hole imaging and for relativistic celestial mechanics, in other words the heart of modern gravitational astronomy \cite{Bailes.al.21}.

Mathematically, Kerr geodesics are described by a system of four ordinary differential equations (ODEs). These ODEs enjoy three remarkable properties:
\begin{enumerate}
    \item they admit a Hamiltonian formulation,
    \item they possess four constants of motion,
    \item these constants are independent and in involution.
\end{enumerate}

Together, these three features imply that Kerr geodesics form an \emph{integrable} Hamiltonian system, in the classical (Arnold-Liouville) sense \cite{Arn,HinFle.08}. Our work asks whether this property remains when one departs from the test-body (geodesic) picture and considers instead an \emph{extended-body} orbiting a Kerr black hole. Our main motivation for tackling this problem is the following.

The aforementioned integrability underlies most of the machinery developed for modeling extreme mass-ratio inspirals (EMRIs), among the prime sources for the future space-based Laser Interferometer Space Antenna (LISA) \cite{LISAproposal,LISA:2024hlh,LISAWHITE.23}. Integrability implies the existence of action-angle variables and of well-defined fundamental frequencies \cite{Schm.02}. These, in turn, are what make the two-timescale expansion possible \cite{HinFle.08,MiPo.21,Po.al2.11} and what allow waveforms to be partly generated from precomputed grids rather than from direct orbital integration \cite{MiPo.21, PoWa.21, Mathews:2025nyb, Lewis:2025ydo}. Building gravitational waveforms for generic bound orbits (eccentric and inclined), is tractable only because four constants confine the motion to invariant tori on which it is quasi-periodic. In fact, it is an interesting exercise to imagine what EMRI waveform modelling would look like without enough constants of motion. 

The geodesic approximation is not enough, however, for the future of gravitational astronomy. Extracting EMRI parameters at the precision LISA requires means controlling the waveform phase to second post-adiabatic (2PA) order \cite{HinFle.08,PoWa.21}, which in turn requires including effects that enter at first order in the mass ratio beyond the leading dissipative one. Among these is the secondary's spin, whose leading coupling to the background curvature contributes precisely at 2PA \cite{DruHug.I.22,DruHug.II.22,Skoupy:2026ewu,Piovano:2024yks,MaPoWa.22,Mathews:2025txc,MiPo.21, PoWa.21,Lewis:2025ydo}. Remarkably, adding spin brings new degrees of freedom and yet it does not destroy the integrable structure that made the geodesic problem tractable \cite{WitzHJ.19,Ra.Iso.IntegO1.26}. At linear order in spin, energy, angular momentum and Carter constant all admit spin corrections that keep them conserved, and a new constant of motion, the R\"udiger invariant \cite{Rudiger.I.81,Rudiger.II.83}. The entire apparatus (two-timescale expansion, adiabatic inspiral, fundamental frequencies, etc) thus carries over. 

All of the above rests on Kerr geodesics or linear-in-spin dynamics, modeling a test, structureless, point-like object orbiting a black hole. Real objects, however, do not follow this ideal. Many conservative physical phenomena render the geodesic or linear-in-spin approximation insufficient: environmental drag \cite{Lui:2026uai,Copparoni:2025vty}, third body resonances \cite{Bonga:2019ycj}, beyond-GR effects \cite{Zi:2026zpw}. Even setting all of these aside, the dynamics are also driven by couplings between the body's multipoles and the background curvature, which at leading order produce quadrupolar forces and torques \cite{Di.15,Ha.12,Ha.15}. Producing waveform models including these effects is slowly becoming an important task of the community \cite{RahBha.23,RahShaPouMat.26}, both for the LISA mission and for other up-coming gravitational observatories, including LISA \cite{LISAproposal, LISA:2024hlh}, TianQin/Taiji \cite{Gong:2021gvw}, DECIGO \cite{Kawamura:2020pcg}, and future ground-based detectors \cite{Punturo.al.10, Hild:2010id,Reitze:2019iox, Evans:2021gyd}.

\subsection{Quadrupolar integrability}

The question that we ask in this work is the following. If the ODEs governing the particle are not the geodesic (or linear-in-spin) equations, but include the next-order (quadrupolar) effects, what becomes of the special \emph{integrability} feature?  More specifically:
\begin{enumerate}
    \item are the quadrupolar ODEs Hamiltonian?
    \item do they admit constants of motion?
    \item are these constants independent and in involution? 
\end{enumerate}

Partial answers to these interrogations already exist, and depend on multipolar order. At linear order in the body's spin, the dynamics remains Hamiltonian, and the dynamical mass, energy, and angular momentum all admit spin corrections that render them conserved; so does the Carter constant, through the R\"udiger invariants \cite{Di.74,Rudiger.I.81,Rudiger.II.83,ComDru.22,Ra.Iso.IntegO1.26}. At quadratic order in spin, the spin-induced quadrupole enters, controlled by a deformability parameter $\kappa$, and the situation changes: a deformed Carter constant still exists, but only if $\kappa$ takes its black-hole value $\kappa=1$. This is true in a Kerr background \cite{ComDruVin.23,Ra.CQG.24} and, more generally, in spacetimes endowed with the Killing--Yano symmetry \cite{Ra.Iso.Dru.IntegO2.26,deFiVi.26}. 

Another quadrupolar effect is commonly seen to appear in the literature: the \emph{tidally-induced} quadrupole, which describes a (not necessarily spinning) body deformed by, and responding adiabatically to, the curvature of the background through which it moves. For most celestial objects, spin-induced deformations largely dominate over tidal ones. For the Earth, both effects differ by a factor of $\simeq10^5$: due to the Earth's spin, the equatorial radius is about 20~km larger than the polar one, whereas crustal tides induced by celestial objects deform the Earth by approximately 20--30~cm \cite{Sun.al.23}. Like the spin-induced one, the tidal quadrupole strength is set by the body's internal structure through coupling parameters, essentially the electric- and magnetic-type Love numbers \cite{Hi.08,FlHi.08,DaNa2.09}. Several studies have already looked at the dynamics of particles subject to this kind of tidally-induced quadrupole: 
Refs.~\cite{StPu.12,Bi.al.12,Endlich:2015mke} used action/effective-field-theory approach to motivate the quadrupole; 
Ref.~\cite{Bini:2014xyr} also studied the role of a generic (unspecified) quadrupole tensor in the equatorial plane of a Kerr black hole;
Ref.~\cite{Chen.al.19} computed waveforms for equatorial orbits in the near-horizon region of a near-extremal Kerr black hole; 
Refs.~\cite{Hen.al.20,Hen.al2.20,Hen.al3.20} derived the leading order tidal-quadrupole (and octupole) contributions to the post-newtonian Hamiltonian, equations of motion and fluxes. More recently,
Ref.~\cite{RahShaPouMat.26} computed gravitational fluxes for circular, equatorial orbits in Kerr, including both spin-induced and tidally-induced quadrupole.  

\subsection{This work}

In this paper we switch off the body's spin and consider this tidal quadrupole alone, asking whether the three properties satisfied by Kerr geodesics still hold, to leading order in tidal effects. We find that the tidal dynamics remain Hamiltonian, and energy, angular momentum, and a tidally-corrected dynamical mass remain conserved. However, the Carter constant does not: \emph{for generic tidal couplings and generic Kerr spin, it admits no tidal-corrected deformation that is polynomial in the momenta}. The leading-order tidal dynamics in Kerr is therefore not integrable.

The non-existence proof itself proceeds by reducing a completely general polynomial Ansatz, using the symmetries of the problem, to a unique three-function form, and showing that the resulting overdetermined PDE system violates its own integrability conditions, except in the Schwarzschild limit $a=0$, where integrability is preserved by spherical symmetry. Our method can easily generalize to other quadrupole models (beyond tides) and other spacetime backgrounds (beyond Kerr). Along the way, we obtain other results of independent interest. For example, we derive closed-form expressions for the electric and magnetic tidal scalars $(\mcE^2,\mcB^2)$ valid in spacetimes possessing a KY tensor: remarkably, their entire momentum dependence is through the geodesic Carter constant, in the form of a quadratic polynomial with coefficients depending only on the Weyl scalar. We believe these formulae to be new. 

Having established that no tidally-deformed Carter constant exists, we turn to the dynamical consequences, which we exhibit numerically. Three complementary diagnostics, applied to the tidally-perturbed dynamics in Kerr, show the expected signature of non-integrability. Poincar\'e sections reveal the destruction of the invariant tori, with surviving KAM curves, resonant island chains, and a connected chaotic sea. Lyapunov exponents confirm the exponential divergence of neighboring orbits. An escape-time map, recording whether and when an orbit plunges into the black hole as a function of its initial data, exhibits a fractal dependence on initial conditions, scale-invariant over the two decades of resolution we probe. Because our equations of motion are truncated at linear order in the tidal coupling $\epsilon$, and because generic $O(\epsilon^2)$ terms would destroy tori regardless, we have quantitatively confirmed that the observed effects are genuinely of first order. 

In short, the analytical calculations and result of Sec.~\ref{sec:integrability} proves the non-integrability, the numerical diagnostics of Sec.~\ref{sec:chaos} confirm and illustrate its dynamical consequences, and the scaling analysis discussed in \ref{ssec:order} confirms that the latter is sourced by leading-order tidal effects, not sub-leading or numerical artifacts.

\subsection{Organization of the paper}

We start in Sec.~\ref{sec:tidalEOM} with a review of the dynamics of a non-spinning particle endowed with a tidally-induced quadrupole. Building on it, we formulate in Sec.~\ref{sec:Ham} the equations of motion as a Hamiltonian system on an 8-dimensional (8D) phase space. In Sec.~\ref{sec:tidalscalars}, the tidal part of this Hamiltonian is simplified using the Killing symmetries and the algebraic specialty of Kerr: we show that it can be rewritten solely in terms of the Weyl scalar, the particle's dynamical mass and the geodesic Carter constant. In Sec.~\ref{sec:Carterlike}, we show that a tidally-corrected, Carter-like constant of motion cannot exist. Lastly, in Sec.~\ref{sec:chaos}, we present our numerical diagnostics confirming the analytical result: Poincar\'e sections, Lyapunov-exponent computations and escape-time maps. All reveal expected features of the phase space's geometry for non-integrable systems. We close in Sec.~\ref{sec:discussion} with a summary of our results and their consequences, a check that the numerically observed chaos is a genuine first-order effect in the tidal coupling (Sec.~\ref{ssec:order}), and discuss implications for EMRI modelling and potential future works.

Throughout, we use geometric units $G=c=1$ and the metric signature $(-,+,+,+)$. Lowercase Latin indices $a,b,c,\ldots$ are abstract indices, Greek indices $\alpha,\beta,\ldots$ denote tensor components in a natural basis, and Latin indices $i,j,\ldots$ run over the Boyer-Lindquist coordinates $(r,\theta)$, used only in Sec.~\ref{sec:Carterlike}. Our conventions for differential geometry follow Wald \cite{Wald}, in particular the Riemann tensor satisfies $2\nabla_{[a}\nabla_{b]}\omega_c=R_{abc}^{\phantom{abc}d}\omega_d$ for any $\omega_c$.

%%%%%%%%%%%%%%%%%%%%%%%%%%%%%%%%%%%%%%%%%%
\section{Evolution equations with tides}
\label{sec:tidalEOM}
%%%%%%%%%%%%%%%%%%%%%%%%%%%%%%%%%%%%%%%%%%

In this section, we consider a non-spinning test particle subject to a tidally-induced quadrupole.
The quadrupole model that we chose is that used in prior studies on quadrupolar tidal effects of the secondary object, e.g., \cite{StPu.12,Chen.al.19,Hen.al.20,Hen.al2.20,RahShaPouMat.26}. It assumes that the compact object responds adiabatically and linearly to tidal effects induced by the external metric. Although our goal is to study a test object in a Kerr background, the present section and the next make no assumption on the nature of the metric, aside from the fact that it is fixed (there is no back-reaction from the particle). The framework will be applied to a Kerr background starting in Sec.~\ref{sec:tidalscalars}.

%%%%%%%%%%%%%%%%%%%%%%%%%%%%%%%%%%%
\subsection{Evolution of a spinless particle}
\label{subsec:spinless}
%%%%%%%%%%%%%%%%%%%%%%%%%%%%%%%%%%%%

We begin by a brief review on the motion of test bodies in general relativity. Our starting point is the multipolar Dixon-Harte formalism \cite{Di.64,Di.15,Ha.12,Ha.15}, which describes the evolution of generic stress-energy tensor distributions in a background spacetime. The body is described as a wordline endowed with multiple moments, the first three being (i) the four-momentum vector $p_a$, (ii) the antisymmetric spin tensor $S^{ab}$ and (iii) the quadrupole tensor $J^{abcd}$, whose algebraic symmetries match those of the Riemann curvature tensor $R_{abcd}$. These, and all other moments, are defined as hypersurface integrals over the body's stress-energy tensor $T^{ab}$. Under the assumption that the body exerts no back-reaction on the background spacetime, and that its stress-energy tensor satisfies $\nabla_a T^{ab}=0$, the body's momenta are shown to satisfy the Mathisson-Papapetrou-Tulczyjew-Dixon (MPTD) equations \cite{Ma.40,Pa.51,TuTu.62,Di.64,Di.74,Di.15,Ha.12,Ha.15}
\begin{subequations}\label{spinlessMPDeq}
    \begin{align}
        \dot{p}_a &= R_{a b c d} S^{b c} v^d + F_a, \label{dotptidal} \\
        \dot{S}^{ab} &= 2p^{[a}v^{b]} + N^{ab}, \label{torquebal}
    \end{align}
\end{subequations}
where an overdot denotes $v^e\nabla_e$, the covariant derivative along the worldline with respect to a tangent vector $v^a$, and $F_a,N^{ab}$ are the force and torque acting on the particle. The latter can be given explicitly at any multipolar order in terms of the background curvature and the body's quadrupole and higher-order moments. 

The first assumption that we make is that the body has no proper rotation in the sense $S^{ab}=0$. Notice that this does not imply the absence of a torque on the body from Eq.~\eqref{torquebal}, so long as the particle's velocity $v^a$ and four-momentum $p^a$ are misaligned. Reciprocally, a torque-less particle can still have nonzero spin tensor $S^{ab}$, from the same argument, such that being spin-free and torque-free are logically independent in relativistic mechanics, cf. Sec.IV.A in \cite{Ha.20}. 

We define the particle's dynamical mass $\mu$ and the unit momentum $\hat{p}_a$ by
\begin{equation} \label{defmupbar}
    \mu^2=-p_ap^a \quand \hat{p}_a=p_a/\mu.
\end{equation}
Following what is done in the spinning case \cite{Ha.15}, we make the following convenient choice for the tangent vector $v^a$:
\begin{equation} \label{normalizationV}
    v^ap_a=-\mu,
\end{equation}
without loss of generality. Contracting both equations in \eqref{spinlessMPDeq} with $\hat{p}^a$ while using $S^{ab}=0$ and the normalization \eqref{normalizationV} leads to
\begin{align}
    \dot{\mu} &= -\hat{p}^a F_a \label{dotmu},\\
    v^a &= \hat{p}^a - \mu^{- 1} N^{a b} \hat{p}_b. \label{momvelspinless}
\end{align}
Equation \eqref{dotmu} states that the particle's dynamical mass $\mu$ is \textit{not} conserved if the particle is subject to a force with components parallel to the four-momentum. The second equation is the so-called \emph{momentum-velocity relation}, which shows that tangent vector (and thus the worldline) is fixed once the multipoles and the curvature are known.

Although the MPTD equations \eqref{spinlessMPDeq} are invariant under the parametrization of the worldline (i.e., the tangent vector $v^a$), in a perturbative framework we can readily make a statement about the nature of the vector $v^a$ and its associated parameter $\lambda_v$ by taking the norm of \eqref{momvelspinless}. Using the antisymmetry of the torque $N^{ab}$, we find
\begin{equation} \label{normV}
  v^a v_a = - 1 + O([\text{torque}]^2),
\end{equation}
where the explicit form of the second term on the right-hand side is $\mu^{- 2} N^{a b} \hat{p}_b N_{a c} \hat{p}^c$, quadratic in the torque $N_{ab}$. Since $v^a$ is tangent to the worldline, equation \eqref{normV} readily implies
\begin{subequations} \label{utau}
\begin{align}
    v^a &= u^a + O([\text{torque}]^2), \\
    \lambda_v &= \tau + O([\text{torque}]^2),
\end{align}
\end{subequations}
where $u^a$ and $\tau$ are, respectively, the four-velocity and the the proper time of the worldline, such that $u^\alpha=\ud x^\alpha/\ud\tau$, in components. In particular, if one can neglect the remainder in \eqref{normV}, as we will do subsequently, then $v^a$ is nothing but the four-velocity and $\lambda_v$ is the proper time $\tau$.

\subsection{Tidally-induced quadrupole} \label{subsec:tidquad}

We now assume that the force and torque in \eqref{spinlessMPDeq} are given by their quadrupole expression and neglect higher-order multipoles.\footnote{See \cite{Ma.15,HarteRamond.26} and \cite{Amancio:2026mnj} for the octupole and hexadecapole expressions, respectively. Higher-order expressions are difficult to obtain and do not seem to appear in the literature yet.} At quadrupolar order, the force and torque are well-known and given by \cite{Ha.20,Ra.Iso.Dru.IntegO2.26}:
\begin{subequations} \label{defFN}
    \begin{align}
        F_a &= -\tfrac{1}{6}J^{bcde}\nabla_a R_{bcde} \,, \\
        N^{ab} &= \tfrac{4}{3} J^{cde[a} R^{b]}_{\phantom{bi}ecd} \,.
    \end{align}
\end{subequations}
The quadrupole tensor $J^{abcd}$ in \eqref{defFN} is assumed to be tidally-induced, following a model classically used in the literature \cite{Endlich:2015mke,Hen.al.20,Hen.al2.20,Hen.al3.20,RaLe.20,PhDHal.21,Chen.al.19,Bi.al.12,St.15,RaLe.20,PhDHal.21,RahShaPouMat.26}:
\begin{align} \label{Jtidal}
    J^{abcd} &= c_\text{E} \, \hat{p}^{[a}E^{b][c} \hat{p}^{d]} \nonumber\\
    &+ c_\text{B}  \bigl( \, \hat{p}^{[a } B^{ b] g} \varepsilon^{cd}_{\phantom{de} g f} \hat{p}^f + \hat{p}^{[c } B^{ d] g} \varepsilon^{ab}_{\phantom{de} g f} \hat{p}^f \bigr),
\end{align}
where $(c_\text{E},c_\text{B})$ are two dimensionful numerical coefficients. They will be used as parameters with respect to which we linearize, in order to keep only leading-order effects. They are in one-to-one correspondence with the (electric- and magnetic-type) quadrupolar Love numbers of the body which have been the subject of aforementioned studies. We refer to the recent work \cite{RahShaPouMat.26} (in particular Table.~1 and Sec.~B there) for references and explicit values of the tidal coefficients, depending on the secondary's equation of state. Our work makes no assumption on the value of these coefficients. 

The tensors $E_{ab}$ and $B_{ab}$ appearing in equation \eqref{Jtidal} are, respectively, the electric and magnetic parts of the Riemann tensor with respect to the timelike unit vector $\hat{p}_a$. They are symmetric tensors defined by
\begin{equation} \label{elecmag}
        E_{ac} := R_{abcd} \, \hat{p}^b \hat{p}^d \quand B_{ac} := R^\star_{abcd} \, \hat{p}^b \hat{p}^d,
\end{equation}
with $R^\star_{abcd}=\tfrac{1}{2}  \varepsilon_{ab}^{\phantom{ab}ef} R_{cdef}$ denoting the (right) Hodge dual of the Riemann tensor.\footnote{It is also possible to define the left Hodge dual $^\star \!R_{abcd}$ by contracting the Riemann tensor on its first two indices. The difference between left and right duals satisfies $^\star \!R_{abcd}-R^\star_{abcd}=\frac{1}{2}\varepsilon_{abcd}R+2\varepsilon_{cde[a}R^e_{\ph b]}$, which vanishes for vacuum and Einstein ($\Lambda$-vacuum) spacetimes \cite{ComDruVin.23,Ra.Iso.Dru.IntegO2.26}.} Contracting equation \eqref{Jtidal} with $R_{abcd}$ and using \eqref{elecmag} leads to the scalar expression
\begin{equation} \label{JR}
    J^{abcd}R_{abcd}=- c_\text{E} \mcE^2 - 4c_\text{B} \mcB^2,
\end{equation}
where the so-called tidal invariants are given by \cite{Do.al.15,Bini.al.18}
\begin{equation} \label{E2B2}
   \mcE^2 = E_{ab}E^{ab} \quand \mcB^2 = B_{ab}B^{ab}.
\end{equation}
The force and torque induced by the tidal quadrupole are then obtained by combining \eqref{defFN} and \eqref{Jtidal}. One finds
\begin{subequations} \label{tidalFN}
\begin{align}
  F_e &= - \frac{\cE}{6} \hat{p}^a E^{b c} \hat{p}^d \nabla_e R_{a b c d} - \frac{2 \cB}{3} \hat{p}^a B^{b c} \hat{p}^d \nabla_e R^{\star}_{a b c d} \\
  N^{a b} &= \frac{\cE}{3} \hat{p}^c E^{d e} R_{c d e}^{\phantom{c d e} [a} \hat{p}^{b]} + \frac{4 \cB}{3} \hat{p}^c B^{d e} R_{c d e}^{\star \phantom{d e} [a} \hat{p}^{b]},
\end{align}
\end{subequations}

\subsection{Conserved mass}

The identities established above show that equation \eqref{dotmu} implies that the particle's dynamical mass $\mu$ is \emph{not} conserved. In fact, combining \eqref{dotmu} with \eqref{defFN} and $\hp^a=u^a+O(\text{tidal})$, one finds
\begin{equation} \label{dotmu2}
    \dot{\mu}= \frac{1}{6}J^{abcd}\dot{R}_{abcd}.
\end{equation}
However, because of the particular form of the tidal quadrupole \eqref{Jtidal}, the right-hand side in \eqref{dotmu2} can be turned into the following total derivative
\begin{equation} \label{dotmu3}
    J^{abcd}\dot{R}_{abcd} = \frac{1}{2}  \frac{\text{D}}{\ud \tau} \left(J^{abcd}R_{abcd}\right) + O(\text{tidal}^2).
\end{equation}
Comparing \eqref{dotmu2} and \eqref{dotmu3} directly leads to the existence of a conserved mass $\muT$ defined by 
\begin{equation} \label{muT}
    \muT := \mu-\frac{1}{12}J^{abcd}R_{abcd},
\end{equation}
which is conserved in the sense that $\dot{\mu}_\text{T}=O(\text{tidal}^2)$, i.e., at leading order in the quadrupolar tidal effects.\footnote{In the effective field theory and worldine action formalisms, this constant quantity coincides with the value of the Lagrangian taken along solutions to Lagrange's equations \cite{Chen.al.19,Hen.al.20,Hen.al2.20}.} This quantity, which we will call \emph{the} conserved mass, will be crucial for the Hamiltonian formulation in section \ref{sec:Ham}. Note the factor $1/12$ in \eqref{muT}, different from the $1/6$ of the analogous conserved mass quantity in the spinning case (see Eq.~(1.16) in the companion work \cite{Ra.Iso.Dru.IntegO2.26} on spin-induced quadrupole). This implies that no (approximately) conserved mass can be constructed for an arbitrary quadrupole: the pre-factor of $J^{abcd}R_{abcd}$ clearly depends on the explicit form of $J^{abcd}$.

\subsection{Equations of motion}

Gathering all earlier results, we can now write the differential equations governing the eight unknown functions in the problem, namely $\tau\mapsto(x^\alpha(\tau),p_\alpha(\tau))$. They are obtained by combining the two covariant evolution equations \eqref{dotptidal} and \eqref{momvelspinless}, with equations \eqref{utau} and \eqref{tidalFN}, and extracting the components. The final ODEs read
\begin{subequations} \label{tidalODES}
    \begin{align}
        \frac{\ud x^{\alpha}}{\ud \tau} & = \hat{p}^{\alpha} - \frac{\cE}{3 \mu} \left( E^{\delta \varepsilon} R^{\phantom{abc}\alpha}_{\varepsilon \gamma  \delta} \hat{p}^{\gamma} + \mcE^2 \hat{p}^{\alpha} \right) \label{velocityODE}\\
       &\quad - \frac{4\cB}{3 \mu} \left( B^{\delta \varepsilon} R^{\star\phantom{ab\,}\alpha}_{\varepsilon \gamma  \delta} \hat{p}^{\gamma} + \mcB^2 \hat{p}^{\alpha} \right), \nonumber \\
        \frac{\ud p_{\alpha}}{\ud \tau} & = \Gamma_{\gamma\alpha}^{\beta} p_{\beta} \hat{p}^{\gamma} + \frac{\cE}{12} \nabla_{\alpha} \mcE^2 + \frac{\cB}{3} \nabla_{\alpha} \mcB^2.
    \end{align}
\end{subequations}
We note that the electric and magnetic contributions to the right-hand side of the velocity equation \eqref{velocityODE} are actually identical in vacuum type-D spacetimes, as a consequence of the identity
\begin{subequations}
\begin{align}\label{weirdID}
    R_{\alpha \beta \gamma \delta} E^{\alpha \gamma} \hat{p}^{\delta} + \mathcal{E}^2 \hat{p}_{\beta} = R^\star_{\alpha \beta \gamma \delta} B^{\alpha \gamma} \hat{p}^{\delta} +\mathcal{B}^2 \hat{p}_{\beta} 
\end{align}
\end{subequations}
which can be established using the tools presented in section \ref{ssec:nullbiv} below. We also note that the mass $\mu$ appearing in the right-hand side of \eqref{velocityODE} can be replaced by the conserved mass $\muT$, since the difference would lead to $O(\text{tidal}^2)$ effects which we neglect.

%%%%%%%%%%%%%%%%%%%%%%%%%%%%%%%%%%%%%%%%%%
\section{Hamiltonian formulation}
\label{sec:Ham}
%%%%%%%%%%%%%%%%%%%%%%%%%%%%%%%%%%%%%%%%%%

Our goal is now to re-formulate the ODE system \eqref{tidalODES} as a Hamiltonian system on an 8-dimensional (8D) phase space endowed with canonical coordinates $(x^\alpha,p_\beta)$. Two assumptions enter here: that the system be Hamiltonian, and that these coordinates be canonical. Neither is automatic; some systems admit no Hamiltonian form even in non-canonical coordinates.\footnote{For instance $(\dot x,\dot y)=(y,-x-x^3)$ is Hamiltonian with canonical $(x,y)$, but adding a single term to $\dot x$ can change this entirely: $(\dot x,\dot y)=(y+yx^2,-x-x^3)$ is Hamiltonian only if $(x,y)$ are \emph{not} assumed to be canonical, and $(\dot x,\dot y)=(y+x,-x-x^3)$ \emph{cannot} be Hamiltonian at all, regardless of the assumption on coordinates.} In our case the assumption is justified a posteriori: we exhibit the Hamiltonian explicitly and verify that it generates \eqref{tidalODES}.

Naturally, we want the Hamiltonian generating \eqref{tidalODES} to be in the perturbative form
\begin{equation} \label{Htot}
H(x^\alpha,p_\beta) = \Hgeo(x^\alpha,p_\beta) + \Htidal(x^\alpha,p_\beta),
\end{equation}
where $\Hgeo$ generates the geodesic terms and $\Htidal$ accounts for the tidal corrections. We discuss each piece below.

\subsection{Geodesic Hamiltonian} \label{ssec:geoham}

A Hamiltonian that generates geodesics in a given background manifold is well-known, and precedes the advent of general relativity altogether \cite{Arn.66}. A convenient and covariant construction is the following. Let $\mcE$ be a 4D manifold covered with coordinates $x^\alpha$, and $(\partial_\alpha)^a$ be the associated natural bector basis. Consider then the components $p_\alpha$ of the four-momentum in that basis, $p_\alpha := p_a (\partial_\alpha)^a$ . Then, the four pairs $(x^\alpha,p_\alpha)_{\alpha=0,\ldots,3}$ can be used as canonical coordinates covering an 8D phase space $\mcM$, and the (flow of the) following Hamiltonian \cite{Schm.02,HinFle.08}
\begin{equation} \label{Hgeo}
    \Hgeo(x^\alpha,p_\alpha) := \frac{1}{2}g^{\alpha\beta} p_\alpha p_\beta,
\end{equation}
generates the geodesics of $\mcM$ through Hamilton's equations. Here, the metric coefficients $g^{\alpha\beta}$ depend only on the coordinates $x^\alpha$. The Hamiltonian \eqref{Hgeo} is linked to the particle's mass \eqref{defmupbar} via
\begin{equation} \label{H=mu}
    \Hgeo = - \mu^2/2,
\end{equation}
a relation that holds along any geodesic (in spacetime) or, equivalently, any solution to Hamilton's equations (in phase space). On $\mcM$, the usual canonical expression for the Poisson bracket $\{,\}:\mcM\rightarrow\RR$ applies, i.e., for any two functions $F,G$ on $\mcM$, one has 
\begin{equation} \label{PB}
    \{F,G\} = \sum_{\alpha=0\ldots,3} \left(\frac{\partial F}{\partial x^\alpha}\frac{\partial G}{\partial p_\alpha} - \frac{\partial G}{\partial x^\alpha}\frac{\partial F}{\partial p_\alpha} \right).
\end{equation}

The geodesic Hamiltonian \eqref{Hgeo} is said to be \emph{integrable} when it possesses four first integrals in involution, i.e., when there exists four functions $(C_\alpha)_{\alpha=0,\ldots,3}$ on $\mcM$ such that (i) $\{C_\alpha,H\}=0$ for all $\alpha\in\{0,\ldots,3\}$, and (ii) $\{C_\alpha,C_\beta\}=0$ for all $(\alpha,\beta)\in\{0,\ldots,3\}^2$. 

In the Kerr spacetime covered with Boyer-Lindquist coordinates $x^\alpha=(t,r,\theta,\phi)$, one can take $C_0=H$ (the Hamiltonian itself), $(C_1,C_2)=(-p_t,p_\phi)$ (the energy $E=-p_t$ and angular momentum $L_z=p_\phi$ associated to spacetime isometries), and $C_3=\Qgeo$, the eponymous Carter constant \cite{Carter.68}, given by
\begin{equation} \label{QgeoKilling}
    \Qgeo(x^\alpha,p_\alpha) = K^{\alpha\beta}p_\alpha p_\beta ,
\end{equation}
where the $K^{\alpha\beta}$ are the coefficients of the symmetric Killing--St\"ackel (KS) tensor $K^{ab}$ of the Kerr spacetime. From a phase space perspective, the coefficients $K^{\alpha\beta}$ in \eqref{QgeoKilling} depend only on the variables $x^\alpha$. In Kerr, with our conventions for Killing tensors, the Carter constant satisfies
\begin{align}\label{QgeoKerr}
    -\Qgeo^{\text{Kerr}} &= (p_\phi+a p_t)^2 + p_\theta^2 \nonumber\\ & \phantom{=} + \bigl( a^2(\mu^2-p_t^2)+p_\phi^2 \csc^2\theta \bigr) \cos^2\theta ,
\end{align}
where $\mu^2=-2\Hgeo(x^\alpha,p_\beta)$, cf.\ equations \eqref{Hgeo} and \eqref{H=mu}, and $a\in[0,M]$ is the Kerr spin parameter. That $\Qgeo$ is a constant of motion for $\Hgeo$ means, in the Poisson-bracket language, that $\{Q_0,H_0\}=0$. This can either be computed explicitely in Boyer-Lindquist coordinates using \eqref{PB}, or using a covariant (spacetime coordinate-independent) calculation as follows (we detail it as a prototypical example of calculations to come):
\begin{align}
    \{\Qgeo,\Hgeo\} &= p^\beta\{ p_\beta,K^{\gamma\delta}\}p_\gamma p_\delta \nonumber+ p_\alpha p_\beta\{g^{\alpha\beta},p_\gamma \}K^{\gamma\delta}p_\delta \nonumber\\
    &= -p^\beta( \partial_\beta K^{\gamma\delta})p_\gamma p_\delta \nonumber - 2p_\alpha p_\beta (\Gamma_{\gamma\delta}^{\alpha}g^{\beta\delta})K^{\gamma\lambda}p_\lambda \nonumber\\
    &= -p_\alpha p_\beta p_\gamma \nabla^\alpha K^{\beta\gamma} \nonumber\\
    &=0, \nonumber
\end{align}
where, line by line, we have used respectively: the definitions of $(\Qgeo,\Hgeo)$ and the bi-linearity of the Poisson brackets; the Leibniz rule and the identity $\partial_\gamma g^{\alpha\beta} = -2\Gamma_{\gamma\delta}^{(\alpha}g^{\beta)\delta}$ (a consequence of $\nabla_ag_{bc}=0)$; the definition of the covariant derivative; and the defining equation $\nabla_{(a}K_{bc)}=0$ satisfied by the KS tensor.

The fact that the geodesic Hamiltonian is integrable in the Kerr spacetime is a remarkable result, and it unlocks a plethora of tools that have no counterpart in generic Hamiltonian systems. Integrability guarantees that the motion is confined to invariant tori in phase space, on which it is quasi-periodic with directly entering the gravitational waves emmited by asymetric binary systems modeled as a particle orbiting a Kerr background. The reader unfamiliar with these constructions will find in Refs.~\cite{Schm.02} and \cite{HinFle.08} two particularly clear and covariant expositions: the former derives the quasi-periodic (action-angle) formulation of Kerr geodesics and the resulting fundamental frequencies explicitly, while the latter develops the two-timescale framework that this structure makes possible, and shows precisely where the frequencies enter the construction of adiabatic and post-adiabatic inspirals \cite{PoWa.21}.

\subsection{Tidal Hamiltonian}

We now look for the Hamiltonian piece $H_1$ that generates the leading-order corrections brought by the tidal quadrupoles. Since there are no new degrees of freedom brought by tidal effects, and that all expressions can be expressed in terms of the sole four-momentum and geometry, the phase space welcoming $H_0$ and $H_1$ can be assumed to stay the same, i.e., 8D and covered with canonical pairs $(x^\alpha,p_\alpha)_{\alpha=0,\ldots,3}$.

By construction, since the tidal effects are time-independent, the full hamiltonian \eqref{Htot} should be conserved along solutions to Hamilton's equations to leading order in tidal effects. Since $H_0=-\mu^2/2$, we are thus looking for a function $H$ of the phase-space variables $(x^\alpha,p_\beta)$ that is conserved and of dimension $[\text{mass}]^2$. A natural candidate is, therefore, the quantity $-\mu_\text{T}^2/2$ built from the conserved mass $\muT$ that was discussed around Eq.~\eqref{muT}. This takes the form
\begin{equation} \label{muT2exp}
    -\frac{\muT^2}{2}=-\frac{\mu^2}{2} +\frac{\mu}{12} J^{abcd}R_{abcd} 
\end{equation}
when expanded to leading order in tidal effects. Following our comments and \eqref{Hgeo}, equation \eqref{muT2exp} can easily be put in correspondence with \eqref{Htot}. With help from equation \eqref{JR}, we then \textit{define} the following Hamiltonian
\begin{equation} \label{Htidal}
\Htidal(x^\alpha,p_\beta) =  -\frac{\mu}{12} \bigl(c_\text{E} \mcE^2 +4c_\text{B} \mcB^2 \bigr) .
\end{equation}
So defined, this Hamiltonian is a function of the phase space variables $(x^\alpha,p_\alpha)$ alone, via the \textit{covariant} formulae \eqref{defmupbar}, \eqref{elecmag} and \eqref{E2B2}. Crucially, the $\mu$ appearing on the right-hand side must be considered a function of $(x^\alpha,p_\alpha)$, not a parameter like $(\cE,\cB)$. This factor \emph{does} contribute to Hamilton's equations, and could very well be written explicitly as $(-g^{\alpha\beta} p_\alpha p_\beta)^{1/2}$ to emphasize its functional dependence.

Inserting \eqref{Hgeo}-\eqref{Htidal} into \eqref{Htot} leads to the following expression for the total (geodesic+tidal)  Hamiltonian:
\begin{equation} \label{Htotnew}
    H = \frac{1}{2} g^{\alpha\beta} p_\alpha p_\beta -\frac{\mu}{12} \bigl(c_\text{E} \mcE^2 +4c_\text{B} \mcB^2 \bigr),
\end{equation}
with \eqref{defmupbar}-\eqref{E2B2} completing the definition for $(\mu,\mcE,\mcB)$. 
Interestingly, one could write this Hamiltonian as $H=\tfrac{1}{2}\tilde{g}^{\alpha\beta}p_\alpha p_\beta$, where $\tilde{g}^{\alpha\beta}$ depends on both $x^\alpha$ and $p_\alpha$, and is given by
$\tilde{g}^{ab} := g^{ab} - \frac{\cE}{6 \mu} E_{cd} R^{acbd} - \frac{2\cB}{3 \mu} B_{cd} R_{\star}^{acbd} .$
It can be interpreted as an effective metric encoding the tidal fields, with respect to which $\muT$ is the ``norm'' of $p_a$ is conserved.

\subsection{Hamilton's equations}

Next, we verify that the Hamilton equations generated by the Hamiltonian \eqref{Htotnew} does indeed reproduce the ODE system \eqref{tidalODES}. This allows us to find the ``time'' parameter associated to this Hamiltonian, which cannot be prescribed independently.\footnote{Indeed, the parameter $\lambda_\text{H}$ associated to a given Hamiltonian is unique. It is the one that enters Hamilton's law of motion $\ud F/\ud \lambda_\text{H}=\{F,H\}$ for any phase space function $F$, and it satisfies $\{\lambda_\text{H},H\}=1$, i.e., $(\lambda_\text{H},H)$ forms a local canonical pair of phase space coordinates.} The following formulae are useful to compute Hamilton's equations 
\begin{subequations} \label{ids}
    \begin{align}
         \frac{\partial \mu}{\partial p_{\mu}} &= - \hat{p}^{\mu} , \quad
         \frac{\partial \mu}{\partial x^{\mu}} = \Gamma^{\alpha}_{\beta \mu}\hat{p}_{\alpha} p^{\beta}, \\
        \frac{\partial \hat{p}_{\nu}}{\partial p_{\mu}} &= \mu^{- 1}h^{\mu}_{\nu} , \quad
        \frac{\partial \hat{p}_{\nu}}{\partial x^{\mu}} = -\Gamma^{\alpha}_{\beta \mu} \hat{p}_{\alpha} \hat{p}^{\beta} \hat{p}_{\nu} , \label{partialhatp}\\
        \frac{\partial \mcE^2}{\partial p_{\mu}} &= \frac{4}{\mu}E^{\alpha \beta}R_{\alpha \gamma \beta \delta} \hat{p}^{\gamma} h^{\mu\delta} , \quad
        \frac{\partial \mcE^2}{\partial x^{\mu}}=\partial_\mu\mcE^2, \\
        \frac{\partial \mcB^2}{\partial p_{\mu}} &= \frac{4}{\mu}B^{\alpha \beta}R^{\star}_{\alpha \gamma \beta \delta} \hat{p}^{\gamma} h^{\mu\delta} , \quad
        \frac{\partial \mcB^2}{\partial x^{\mu}}=\partial_\mu\mcB^2, 
    \end{align}
\end{subequations}
where $h^{a}_b:=\delta^{a}_b+\hp^a \hp_b$ and partial derivatives here only hit objects that depend on $x^\alpha$ explicitly. Using these identities, we can compute Hamilton's canonical equations
\begin{equation}\label{HamEq}
    \frac{\ud x^\alpha}{\ud\lambdaH} = \frac{\partial H}{\partial p_\alpha} \quand \frac{\ud p_\alpha}{\ud\lambdaH} = -\frac{\partial H}{\partial x^\alpha},
\end{equation}
where $\lambdaH$ is the evolution parameter associated to $H$, to be determined. Substituting \eqref{Htotnew} for $H$ and using the identities \eqref{ids}, Hamilton's equations \eqref{HamEq} read
\begin{subequations}\label{ODEsMPD}
    \begin{align}
  \frac{\ud x^\mu}{\ud \lambdaH} &= \muT \hat{p}^{\mu} -\frac{\cE}{3} (E^{\alpha \beta} R^{\phantom{\alpha \gamma \beta} \mu}_{\alpha \gamma \beta}  \hat{p}^{\gamma} + \mcE^2 \hat{p}^{\mu}) \nonumber \\
  & \phantom{=} -\frac{4\cB}{3} (B^{\alpha \beta} R^{\star\phantom{ \gamma \beta} \mu}_{\alpha \gamma \beta}  \hat{p}^{\gamma} + \mcB^2 \hat{p}^{\mu}) , \\
  \frac{\ud p_\mu}{\ud \lambdaH} &= \muT \Gamma^{\alpha}_{\mu\beta}  \hat{p}_{\alpha} p^{\beta} + \frac{\muT}{12} \cE
  \nabla_{\mu} \mcE^2 + \frac{\muT}{3} \cB \nabla_{\mu} \mcB^2 .
    \end{align}
\end{subequations}
Compared to the ODE system \eqref{tidalODES} obtained directly from the Dixon--Harte equations, we find agreement between the two if and only if the Hamiltonian parameter $\lambdaH$ is related to the proper time $\tau$ through $\ud\tau = \muT \,\ud \lambdaH$. This implies that the Hamiltonian \eqref{Htotnew} generates the correct MPTD equations \eqref{tidalODES} with respect to the Hamiltonian ``time'' 
\begin{equation} \label{lambdaH}
   \lambdaH = \frac{\tau}{\muT},
\end{equation}
up to $O(\text{tidal}^2)$. This is consistent with a dimensional analysis, since Hamilton's equations imply that $\lambdaH$ has dimension $[\text{mass}]^{-1}\times[\text{time}]$, and with the geodesic limit, for which we know that $\Hgeo$ generates the geodesic equation with evolution parameter $\tau/\mu$.

\subsection{Summary of the Hamiltonian formulation} \label{ssec:HamSum}

The Hamiltonian formulation of the dynamics a spin-free test particle with a tidally-induced quadrupole is now complete. Like its geodesic counterpart, the formulation is covariant (invariant under spacetime diffeomorphisms) and applies to any background spacetime. We summarize the results below:
\begin{enumerate}[leftmargin=*, align=left, labelindent=0pt, itemindent=0pt, labelsep=0.5em]
    \item the phase space is 8-dimensional and endowed with canonical coordinates $(x^{\alpha},p_\alpha)$;
    \item the Hamiltonian \eqref{Htotnew} generates the MPTD equations with the tidally-induced quadrupole \eqref{Jtidal};
    \item Hamilton's equations are written with respect to the parameter $\tau/\muT$, cf.~\eqref{lambdaH};
    \item $H$ itself is a constant of motion, related to the conserved mass $\muT$ \eqref{muT} via $H=-\muT^2/2$.    
\end{enumerate}

Starting with next section, we specialize the background to the Kerr spacetime and derive a compact closed form for the tidal Hamiltonian \eqref{Htidal}.

%%%%%%%%%%%%%%%%%%%%%%%%%%%%%%%%%%%%%%%%%%
\section{Tidal scalars in Kerr spacetime}
\label{sec:tidalscalars}
%%%%%%%%%%%%%%%%%%%%%%%%%%%%%%%%%%%%%%%%%%

With the Hamiltonian framework complete, we now proceed to rewrite the tidal scalars \eqref{E2B2} that appear in it. These new expressions are more practical to work with, reveal the momentum dependence of the tidal scalars, and make their numerical evaluation more efficient.

\subsection{Null bivector decomposition} \label{ssec:nullbiv}

Our goal is to rewrite the tidal scalar invariants built from the curvature. We thus start with the following convenient decomposition of the Riemann tensor
\begin{equation}\label{Weyl}
  R_{abcd}  =  \,\text{Re} [\Psi (3 Z_{ab} Z_{cd} + 2 g_{a [c} g_{d] b} -
  \ui \varepsilon_{abcd})] ,
\end{equation}
which, although we specialize to Kerr, holds for any vacuum type-D spacetime. In \eqref{Weyl}, $\Psi\in\CC$ is the only non-vanishing Weyl scalar and $Z_{a b}$ is a complex bivector built from an orthonormal null tetrad. We refer to Sec.~III.B of \cite{Ra.Iso.Dru.IntegO2.26} for details on the bivector formalism, which is inspired by Refs.~\cite{Ha.20,ComDruVin.23}. Only the necessary results are presented below. 

The key properties of $Z_{ab}$ that we will need are
\begin{equation}\label{idsZ}
    Z_{a b} Z^{a b} = - 4, \quad Z_{a b} \bar{Z}^{a b} = 0
   \quand Z_{a b}^{\star} = \ui Z_{a b},
\end{equation}
where an overbar denotes complex conjugation and a star denotes the Hodge dual of a bivector, defined by $Z_{ab}^{\star}  =  \frac{1}{2} \varepsilon_{ab}^{\phantom{ab}cd} Z_{cd}$. Note that the first and third equations above imply $\bar{Z}_{a b} \bar{Z}^{a b} = -4$ and $\bar{Z}^{\star}_{a b} = - \ui \bar{Z}_{a b}$, respectively. The tensor $Z_{a b}$ can be used to construct both the metric $g_{ab}$ and the Killing--Yano tensor $f_{ab}$ of the Kerr spacetime, as follows:
\begin{equation} \label{deff}
  g_{ab} = Z_{ac}Z^{c}_{\ph b} \quand f_{ab}  =  \,\text{Re} [C Z_{ab}] ,
\end{equation}
where $C$ is a complex scalar related, via the Bianchi identity, to the Weyl scalar $\Psi$ as follows:
\begin{equation} \label{PsiC3}
  \Psi C^3  =  \ui M
\end{equation}
where $M$ is the Kerr mass parameter (see App.~E in \cite{Ra.Iso.Dru.IntegO2.26} for details).

The Killing-Yano (KY) tensor $f_{ab}$ of the Kerr spacetime is at the root of all its symmetries (e.g., the existence of the two Killing vectors and a KS tensor) and geometric properties (e.g., its Petrov type-D nature). Many of them are explored in our companion works on integrability for spinning particles \cite{Ra.Iso.IntegO1.26,Ra.Iso.Dru.IntegO2.26}, where it plays the central role. We refer to Sec.~III there for more about KY tensors, and will only focus on what is necessary for our development here. In particular, crucially, the KS tensor $K_{ab}$ is given as the ``square'' of the KY tensor:
\begin{equation}
    K_{ab} = f_{ac}f^{c}_{\ph b}. 
\end{equation}

Let us now contract equation \eqref{Weyl} with $\hat{p}^b \hat{p}^d$ to construct the electric part \eqref{elecmag} of the curvature. Since $\hat{p}^b \hat{p}^d$ is real-valued and $\varepsilon_{abcd}$ is totally antisymmetric, we obtain
\begin{equation}\label{defz}
  E_{ab}  =  \,\text{Re} [\Psi (3 z_a z_b - h_{ab})] \quad \text{where}
  \quad z_a := Z_{ab} \hat{p}^b,
\end{equation}
and $h_{ab} := g_{ab} + \hat{p}_a \hat{p}_b$ projects onto the subspace orthogonal to $\hat{p}^a$. In addition, the vector $z_a$ defined here satisfies several identities that require the introduction of the following tensor:
\begin{equation}\label{defcalZ}
    \mathcal{Z}_{ac} := Z_{ab} \bar{Z}^b_{\ph c},
\end{equation}
which is both real-valued and symmetric, like the metric tensor $g_{ab}$. In fact, both are involved in the decomposition of the symmetric KS tensor $K_{a b}$, as follows:
\begin{equation} \label{decompK}
  K_{ab}  =  \frac{1}{2} \,\text{Re} [C^2] g_{ab} + \frac{1}{2} |C|^2 \mathcal{Z}_{ab},
\end{equation}
where $C$ was defined through \eqref{PsiC3}. We refer to \cite{Ra.Iso.Dru.IntegO2.26} for a proof and discussion of this formula.

\subsection{Formulae for the tidal scalars}

We are now ready to construct the formulae for $(\mathcal{E}^2,\mathcal{B}^2)$. Using all aforementioned results, we first note that the vector $z^a$ appearing in \eqref{defz} satisfies:
\begin{equation}\label{idsz}
z_a z^a = 1, \quad z_a \hat{p}^a = 0 \quand z_a \bar{z}^a = - \mathcal{Z}_{a b} \hat{p}^a \hat{p}^b.
\end{equation}
We can then contract $E_{a c}$ in \eqref{defz} with itself over both indices to produce the tidal scalar $\mathcal{E}^2$, cf.~\eqref{E2B2}. Using identities \eqref{idsz}, we find
\begin{equation}\label{Einterm}
    \mathcal{E}^2 = 3\,\text{Re}[\Psi^2] + \frac{3}{2}|\Psi|^2\bigl[3 (\mathcal{Z}_{ab}\hat{p}^a\hat{p}^b)^2-1 \bigr].
\end{equation}
On the other hand, contracting \eqref{decompK} with $\hat{p}^a \hat{p}^b$ gives a relation between the (normalized) Carter constant $\hat{Q}_0 := \Qgeo / \mu^2$ and other scalar fields, namely
\begin{equation} \label{Qhat}
  2 \hat{Q}_0  =  - \,\text{Re} [C^2] + |C|^2 (\mathcal{Z}_{ab} \hat{p}^a \hat{p}^b) .
\end{equation}
Combining equations \eqref{Einterm} and \eqref{Qhat} then gives our final formulae for $\mathcal{E}^2$ (and $\mathcal{B}^2$, following the same recipe). They read
\begin{subequations} \label{finalEB}
    \begin{align}
        \mathcal{E}^2  &=  3 \,\text{Re} [\Psi^2] + \Gamma\bigl( 3 ( 2 \hat{Q}_0 + \,\text{Re} [C^2] )^2 - |C|^4\bigr),\\
        \mathcal{B}^2  &= \mathcal{E}^2 + 6 \,\text{Re} [\Psi^2], \quad \text{with}\quad \Gamma = \frac{3|\Psi|^2}{2|C|^4}.
    \end{align}
\end{subequations}
Several comments can be made on the result \eqref{finalEB}. First, the dependence on either $\Psi$ or $C$ can be dropped using \eqref{PsiC3}. Second, these equations express the tidal scalars $(\mathcal{E}^2,\mathcal{B}^2)$ solely via the complex scalar $\Psi$ (or $C$) and the normalized Carter constant $\hat{Q}_0$. Out of these three scalars, only $\hat{Q}_0$ depends on the momenta. While $\mathcal{E}^2 = R_{a e b f} R_{c \ph d \ph}^{\ph e \ph f} \hat{p}^a \hat{p}^b \hat{p}^c \hat{p}^d$ is a quartic polynomial in the momenta, its momentum dependence is entirely encoded into the normalized Carter constant. Third, we believe that these formulae are new, and we have checked them numerically in Kerr. These calculations and other checks can be found in the attached Mathematical Notebook \cite{MMA_IntegTidal}.

Expanding equation \eqref{finalEB} in powers of $\hat{Q}_0$ and gathering the coefficients gives a formula in the form
\begin{equation}\label{E2simp}
  \mathcal{E}^2 (x^{\alpha}, p_{\beta})  = \sum_{K=0}^2 \EE_K(\Psi) \,\hat{Q}_0^K ,
\end{equation}
where the scalars $\EE_0, \EE_1, \EE_2$ are independent of $p_{\alpha}$, and depend on $x^{\alpha}$ only through $\Psi$. Their expressions are
\begin{subequations}\label{E0E1E2}
    \begin{align}
    \EE_0  &=  3 \,\text{Re} [\Psi^2] + \Gamma \!\left( 3\,\text{Re} [C^2]^2 - |C|^4 \right), \\
    \EE_1 &= 12\,\Gamma\,\text{Re} [C^2], \quand \EE_2 = 12\,\Gamma, \label{finalEB2}
  \end{align}
\end{subequations}
with $C=(\ui M/\Psi)^{1/3}$, cf. \eqref{PsiC3}, and $\Gamma$ defined in \eqref{finalEB2}. The same recipe applies to the magnetic tidal scalar:
\begin{equation}\label{B2simp}
  \mathcal{B}^2(x^{\alpha}, p_{\beta}) = \sum_{K=0}^2 \BB_K(\Psi) \,\hat{Q}_0^K,
\end{equation}
where $(\BB_0, \BB_1, \BB_2) = \left( \EE_0 + 6\,\text{Re} [\Psi^2], \EE_1, \EE_2 \right)$. As a consistency check, we observe that
\begin{equation} \label{Kretschmann}
  \mathcal{E}^2 - \mathcal{B}^2  =  - 6 \,\text{Re} [\Psi^2]
\end{equation}
is independent of the momentum, which is expected since $\mathcal{E}^2 - \mathcal{B}^2$ equals (one eighth of) the Kretschmann scalar $R_{a b c d} R^{a b c d}$ \cite{GourgoulhonBH}. Indeed, that can also be checked directly using equations \eqref{Weyl} and \eqref{idsz}. 

\subsection{Tidal Hamiltonian in closed form}

We can now insert the new expressions of the tidal invariants \eqref{E2simp} and \eqref{B2simp} into the Hamiltonian \eqref{Htidal} to get the compact form
\begin{equation} \label{Htidalclosed}
  \Htidal = -\frac{\mu}{12} \sum_{K=0}^{2} f_K \, \hat{Q}_0^K ,
  \quad
  f_K := \cE\, \EE_K + 4\cB\, \BB_K .
\end{equation}
This is the form that will be used in the remainder of the paper. It makes manifest a key structural feature: \emph{the entire momentum dependence of the tidal perturbation is via the two geodesic invariants $\mu$ and $Q_0$}, the position dependence being confined to the three scalar functions $f_K$, themselves functions of the Weyl scalar $\Psi$ only. Note also, from \eqref{E0E1E2} and the relation between $\BB_K$ and $\EE_K$, that $\BB_K=\EE_K$ for $K=1,2$, so that
\begin{equation}\label{f12prop}
  f_1 = (\cE+4\cB)\,\EE_1 \quand f_2 = (\cE+4\cB)\,\EE_2 ,
\end{equation}
while $f_0$ involves the combination $\cE\,\EE_0+4\cB\,\BB_0$ with $\BB_0\neq\EE_0$. This feature will play an important role in the analysis of Sec.~\ref{sec:Carterlike}.

%%%%%%%%%%%%%%%%%%%%%%%%%%%%%%%%%%%%%%%%%%
\section{Carter-like constant with tides}
\label{sec:Carterlike}
%%%%%%%%%%%%%%%%%%%%%%%%%%%%%%%%%%%%%%%%%%

In this section, we try to construct a Carter-like constant that is preserved by the tidal dynamics, to leading order in the tidal effects. More precisely, we look for a phase-space function
\begin{equation}\label{decompQ}
    Q(x^\alpha,p_\beta) := \Qgeo(x^\alpha,p_\beta) + \Qtidal(x^\alpha,p_\beta),
\end{equation}
where $\Qgeo$ is the usual Carter constant \eqref{QgeoKilling} preserved by $H_0$, and $\Qtidal$ is a correction that is linear in tidal effects, which we denote by $\Qtidal = O(\text{tidal})$. By definition, asking that $Q$ be a constant of motion means that
\begin{equation} \label{PBQH}
    \{Q,H\} = O(\text{tidal}^2).
\end{equation}
Inserting the decompositions \eqref{Htot} and \eqref{decompQ} into the above, and using the fact that $\{ \Qgeo,\Hgeo \}=0$ (by definition of the Carter constant) and $\{\Qtidal,\Htidal\}=O(\text{tidal}^2)$, equation \eqref{PBQH} is equivalent to
\begin{equation}\label{cohomo}
    \{\Qtidal, \Hgeo\} = - \{\Qgeo, \Htidal\}.
\end{equation}
This is a set of PDEs known as the \emph{cohomological equation} of the problem: the right-hand side is known, and the left-hand side is linear in the unknown $\Qtidal$. Our aim is to show that this equation has no solution for generic values of the Kerr spin $a$ and the tidal couplings $(\cE,\cB)$. We proceed in three steps: first, we motivate a general Ansatz for $\Qtidal$; second, we use (spacetime-induced) phase-space symmetries to constrain it into a unique form; third, we derive integrability conditions and show that they are violated.

\subsection{Initial Ansatz for $\Qtidal$}

The correction $\Qtidal$ must be a scalar on phase space, constructed covariantly from the spacetime geometry and the particle's 4-momentum $p_a$, the only dynamical variable characterising the particle. Scalars built algebraically from tensor fields and a single vector are necessarily polynomial in that vector \cite{Kolokoltsov.83,BolFom.04,Frolov.17}. This polynomial structure underlies all known constants of motion in general relativistic dynamics: the energy and angular momentum (from Killing vectors \cite{Ha.12,Di.15}), the Carter constant (from the KS tensor \cite{Carter.68}), and their finite-size generalisations at linear and quadratic order in spin (from Killing--Yano tensors \cite{Rudiger.I.81,Rudiger.II.83,ComDru.22,Ra.Iso.IntegO1.26,ComDruVin.23,Ra.Iso.Dru.IntegO2.26,deFiVi.26}).

Combining these considerations, we write
\begin{equation}\label{Ansatz0}
    \Qtidal = \sum_{n \geqslant 0} \mu^{1-n} \, T^{a_1 \ldots a_{n}}(x) \, p_{a_1} \ldots p_{a_{n}},
\end{equation}
where the $T^{a_1 \ldots a_{n}}$ are rank-$n$ symmetric tensor fields on spacetime, while the exponent $1-n$ of $\mu = (-g^{ab}p_ap_b)^{1/2}$ ensures that each term in the sum has the same overall degree in momenta. This degree is fixed by the cohomological equation \eqref{cohomo}: the right-hand side has degree 2 in momenta (since $\Hgeo$ and $\Qgeo$ both have degree 2, and $\Htidal$ has degree 1), while $\{\Hgeo,\Qtidal\}$ has degree $2+d(\Qtidal)-1$, forcing $d(\Qtidal)=1$.

\subsection{Reduction by symmetries} \label{sec:reductionSyms}

We now show that the symmetries of the Kerr spacetime, combined with the algebraic structure of the reduced phase space, collapse the general Ansatz \eqref{Ansatz0} into a highly constrained form depending only on two scalar functions: the Weyl scalar $\Psi$ and the normalized Carter constant $\hat{Q}_0=Q_0/\mu^2$.

\subsubsection{Phase space reduction}

The Kerr background is stationary and axisymmetric, and the tidal perturbation, built from the geometry, inherits both symmetries. In particular, $E = -p_t$ and $L_z = p_\varphi$ are conserved by the full Hamiltonian \eqref{Htotnew}, and may be treated as fixed parameters. The nontrivial dynamics is thus confined to the reduced 4D phase space covered by $(r,\theta,p_r,p_\theta)$, and the tensor fields $T^{a_1\ldots a_n}$ in the Ansatz \eqref{Ansatz0} can be assumed, without loss of generality, to live on the 2D Kerr sub-manifold spanned by coordinates $(r,\theta)$. Absorbing the $\mu^{-n}$ into the momenta $\hat{p}_a=p_a/\mu$ in \eqref{Ansatz0}, the Ansatz now reads
\begin{equation}\label{Ansatz1}
    \Qtidal = \mu \sum_{n \geqslant 0} T^{i_1 \ldots i_{n}}(r,\theta) \, \hp_{i_1} \ldots \hp_{i_{n}},
\end{equation}
where the indices $(i_1,\ldots,i_n)$ run over $(r,\theta)$ only, and the tensors $T^{i_1 \ldots i_n}$ are viewed as functions of $(r,\theta)$. Our convention for $T^{i_1 \ldots i_n}$ with $n=0$ is a scalar field $T$ (no index).

\subsubsection{Finite rank truncation}

Next, we argue that the sum in~\eqref{Ansatz1} must truncate at finite rank, namely rank 4 ($n\leqslant4$). Indeed, let us write $\Htidal$ as $\Htidal=\mu h_1$, where $h_1$ is a quartic polynomial in momentum, cf.\ \eqref{Htidalclosed}. Now, one has
\begin{equation} \label{PBQ0mu}
    \{\Qgeo,\mu\}=-\{\Qgeo,\Hgeo\}/\mu=0,
\end{equation}
where we used the Leibniz rule and $\Hgeo=-\mu^2/2$ in the first equality, and the conservation of Carter's constant in the second. Using \eqref{PBQ0mu}, the right-hand side of \eqref{cohomo} is thus $-\mu\{\Qgeo,h_1\}$, with the latter bracket being quintic in momentum, as $\Qgeo$ and $h_1$ are quadratic and quartic in it, respectively. Matching degrees on both sides of \eqref{cohomo} then forces the polynomial part of $\Qtidal$ to be \emph{at most} quartic in $(p_r,p_\theta)$ as well. Consequently, the Ansatz becomes
\begin{equation}\label{Ansatz2}
    \Qtidal = \mu \sum_{n=0}^4 T^{i_1 \ldots i_{n}}(r,\theta) \, \hp_{i_1} \ldots \hp_{i_{n}},
\end{equation}
with the five tensors $(T,T^i,T^{ij},T^{ijk},T^{ijkl})$ arbitrary.

\subsubsection{Parity and degree constraint}

In Boyer--Lindquist coordinates, both the Kerr metric components $g^{\alpha\beta}$ and the Killing tensor components  $K^{\alpha\beta}$ are block-diagonal: $g^{r\theta}=0=K^{r\theta}$. As a consequence, $\Hgeo$, $\Htidal$, and $\Qgeo$ are each separately even under the two discrete symmetries $p_r\to -p_r$ and $p_\theta\to-p_\theta$. The right-hand side of the cohomological equation \eqref{cohomo}, $\{\Qgeo,\Htidal\}$, is therefore odd in both $p_r$ and $p_\theta$. It follows\footnote{Indeed, suppose $\Qtidal$ has an odd-in-$p_r$ component $\Qtidal^{\text{odd}}$. Then $\{ \Qtidal^{\text{odd}},\Hgeo\}$ is even in $p_r$, but the right-hand side of \eqref{cohomo} is odd. The even part of the equation thus requires $\{\Hgeo, \Qtidal^{\text{odd}}\} = 0$, meaning that $\Qtidal^{\text{odd}}$ would be a geodesic first integral that is odd in $p_r$. No such integral exists in Kerr. Hence $\Qtidal^{\text{odd}} = 0$, and $\Qtidal$ is even in $p_r$. The identical argument applies to $p_\theta$.} that $\Qtidal$, if it exists, must be separately even under both $p_r\to -p_r$ and $p_\theta\to-p_\theta$.

This parity constraint, combined with the finite sum \eqref{Ansatz2}, readily implies that the odd-rank tensors $T^i$ and $T^{ijk}$ must vanish, since these necessarily contribute odd-in-$\hp_i$ terms. Only the three even-rank tensors ($n=0,2,4$) survive, and the Ansatz \eqref{Ansatz2} reduces to
\begin{align}\label{Ansatz3}
    \Qtidal = \mu \bigl( T + T^{ij}\hp_{i} \hp_{j} + T^{ijkl} \hp_{i} \hp_{j} \hp_{k} \hp_{l} \bigr),
\end{align}
with the scalar $T$ and symmetric tensors $(T^{ij},T^{ijkl})$ still unconstrained at this stage.

\subsubsection{Basis for diagonal tensors} \label{sssec:basis}

Let us now focus on the unconstrained symmetric tensor $T^{ij}$ appearing in \eqref{Ansatz3}. It lives in the 2D space spanned by coordinates $(r,\theta)$. As such, it possesses 3 independent components ($T^{rr},T^{r\theta},T^{\theta\theta}$), but the parity constraint established earlier imposes $T^{r\theta}=0$. In other words, $T^{ij}$ is symmetric and diagonal. A convenient basis for such tensors in 2D is that made of the metric and the KS tensor themselves:
\begin{equation} \label{basisTij}
    T^{ij}\in \text{span}[g^{ij},K^{ij}] ,
\end{equation}
which are indeed symmetric and diagonal in the $(r,\theta)$-subspace, but most importantly, linearly independent.

Similar considerations apply to the rank-4 tensor $T^{ijkl}$: it must be diagonal by the parity constraint, and thus possesses 3 independent components. Similarly, a tensor basis is made of the symmetrized products between the metric and KS tensor:
\begin{equation} \label{basisTijkl}
    T^{ijkl}\in \text{span}[\,g^{(ij}g^{kl)},\, g^{(ij}K^{kl)},\,K^{(ij}K^{kl)}\,].
\end{equation}

Let us now expand the tensors $T^{ij},T^{ijkl}$ appearing in \eqref{Ansatz3} onto the bases \eqref{basisTij} and \eqref{basisTijkl}, perform the contraction with the reduced momentum, and use the identities $g^{ij} \hp_i \hp_j=-1$ and $K^{ij} \hp_i \hp_j=\hat{Q}_0$ to obtain the final form of the Ansatz for $\Qtidal$:
\begin{equation}\label{Q1final}
  \Qtidal = \mu\Bigl(A_0(r,\theta) + A_1(r,\theta)\,\hat{Q}_0 + A_2(r,\theta)\,\hat{Q}_0^2\Bigr) \,,
\end{equation}
where the functions $A_0,A_1,A_2$ depend only on $(r,\theta)$. Once again, we emphasize that this is the unique structure forced by the parity and degree constraints, together with the completeness of $(g^{ij}, K^{ij})$ as building blocks for a basis of diagonal tensors in the reduced $(r,\theta)$ space.

Importantly, comparing \eqref{Q1final} with \eqref{Htidalclosed}, one sees that $\Qtidal$ ends up looking exactly like $\Htidal$ itself: the product of $\mu$ with a quadratic-in-$\hat{Q}_0$ polynomial with $(r,\theta)$-dependent coefficients.

\subsection{Integrability conditions} \label{sec:integrability}

We now substitute the form \eqref{Q1final} into the cohomological equation \eqref{cohomo} and derive necessary conditions for the existence of the three unknown functions $A_0, A_1, A_2$ appearing in \eqref{Q1final}.

\subsubsection{Deriving integrability conditions}

Since $\{\Hgeo, \mu\}=0$ and $\{\Hgeo, \hat{Q}_0\}=0$ (the dynamical mass and the normalised Carter constant are geodesic invariants), the left-hand side of the cohomological equation \eqref{cohomo} factorises cleanly at each power of $\hat{Q}_0$:
\begin{equation}\label{LHS}
  \{\Hgeo, \Qtidal\} = \mu \sum_{K=0}^{2} \hp^i (\partial_i A_K)  \, \hat{Q}_0^K ,
\end{equation}
where $i$ is the tensor index running over $\{r,\theta\}$, and $K$ is the summing index, labeling the scalar functions $A_0,A_1,A_2$ of \eqref{Q1final} and the powers of $\hat{Q}_0$.

For the right-hand side of \eqref{cohomo}, we use the closed form \eqref{Htidalclosed} of the tidal Hamiltonian. Since $\{\mu, \Qgeo\}=0$ and $\{\hat{Q}_0, \Qgeo\}=0$, only the position derivatives of $f_K$ contribute to the bracket, giving
\begin{equation}\label{RHS}
  \{\Qgeo, \Htidal\} = \frac{\mu}{6} \sum_{K=0}^{2} \hp_i \,K^{ij}(\partial_j f_K) \, \hat{Q}_0^K ,
\end{equation}
where, again, $K^{ij}$ are the components of the KS tensor $K^{ab}$ in the $(r,\theta)$ space.

Equating \eqref{LHS} and \eqref{RHS} at each power of $\hat{Q}_0$, and matching the coefficients of $\hp_r$ and $\hp_\theta$ independently, yields \emph{three} sets of first-order PDEs (one for each $K=0,1,2$):
\begin{equation}\label{PDEsystem}
  \partial_i A_K = -\frac{1}{6}\,K_{i}^{\phantom{i}j}\,\partial_j f_K.
\end{equation}

For a given $K$, equation \eqref{PDEsystem} is a system of two first-order PDEs (one for $i=r$, one for $i=\theta$) constraining a single unknown scalar function $A_K(r,\theta)$. More specifically, it constrains the two partial derivatives of $A_K(r,\theta)$ in terms of known quantities: $K_{i}^{\phantom{i}j}$ are the components of the KS tensor, and the functions $f_K$ are given by equation \eqref{Htidalclosed} and depend exclusively on $(r,\theta)$ through the Weyl scalar, cf. \eqref{E0E1E2} and \eqref{PsiC3}. Equation \eqref{PDEsystem} is an overdetermined system, and solutions exist if and only if the right-hand side has a special property. Indeed, consider taking a second partial derivative $\partial_\ell$ of \eqref{PDEsystem} and anti-symmetrizing over the two derivative indices: by Schwarz's theorem we must have $\partial_{[\ell}\partial_{i]}A_K=0$. This readily implies the following \emph{integrability conditions} for the three right-hand sides (one for each $K=0,1,2$):
\begin{equation}\label{compatcond}
  \partial_{[r}\bigl(K_{\theta]}^{\phantom{ii}i}\,\partial_i f_K\bigr) = 0.
\end{equation}
These are three differential conditions on the known functions $f_K(r,\theta)$, which depend on the tidal couplings $(\cE,\cB)$ and the Kerr parameters $(a,M)$ through the Weyl scalar $\Psi$ and the KS components $K_{i}^{\phantom{i}j}$. To summarize: without solving the PDE system \eqref{PDEsystem}, the form of the equation already constrains whether it can have a solution. Let us now analyze the constraints \eqref{compatcond}.

\subsubsection{Analysis of the integrability conditions}

A key structural feature of the coefficients $f_K$ follows from the relation $\BB_K = \EE_K$ for $K=1,2$, already noted in \eqref{f12prop}. The $K=1$ and $K=2$ compatibility conditions \eqref{compatcond} are therefore automatically satisfied if $\cE + 4\cB = 0$, and are nontrivial otherwise. Direct evaluation in Boyer--Lindquist coordinates (cf. the attached Mathematica Notebook \cite{MMA_IntegTidal}) shows that \eqref{compatcond} fails for $K=2$ whenever $a\neq 0$, yielding a nonzero scalar proportional to $a$. This eliminates all tidal couplings with $\cE + 4\cB \neq 0$.

It remains to examine the $K=0$ condition when $\cE+4\cB=0$. In this case, $f_1=f_2=0$ and $\Qtidal = \mu\, A_0(r,\theta)$ from \eqref{Q1final}. This has no momentum dependence beyond the overall factor of $\mu$. The tidal Hamiltonian \eqref{Htidalclosed} then reduces to a very simple form
$\Htidal = \frac{\cE}{2}\mu\,\text{Re}[\Psi^2]$,
and the $K=0$ condition \eqref{compatcond} is \emph{not satisfied} (as we computed in the joined Mathematica notebook). Moreover, one can see that it would require $\text{Re}[\Psi^2]$ to be separable as a function of $r$ plus a function of $\theta$. But for any $a\neq 0$, this is not the case: with $\Sigma:=r^2+a^2\cos^2\theta$, one has
\begin{align} \label{RePsi2}
    \text{Re}[\Psi^2] &= \frac{M^2}{\Sigma^6} \Bigl(r^6-a^6 \cos^6\theta \nonumber \\
    & \qquad - \underbrace{15 \,a^2 r^2 ( r^2  - a^2  \cos^2\theta ) \cos^2\theta}_{\text{inseparable cross-terms}}\,  \Bigr).
\end{align}

\subsection{Conclusion}

Our conclusion is that none of the integrability conditions \eqref{compatcond} are satisfied. The PDE system \eqref{PDEsystem} therefore has no solution, and no quantity $Q=\Qgeo+\Qtidal$ can be built such that \eqref{PBQH} holds: the geodesic Carter constant \emph{does not} admit a tidal correction that makes it a new conserved quantity under tidal dynamics.

This result a for tidal-induced quadrupole contrast with a spin-induced one, for which a deformed Carter constant does exist, conditionally on the coupling taking its Kerr black-hole value \cite{ComDruVin.23,Ra.Iso.Dru.IntegO2.26}. We discuss it further in Sec.~\ref{sec:discussion}, where we argue that integrability at quadrupolar order singles out binary black holes.

\subsection{The Schwarzschild limit}

In the Schwarzschild limit ($a=0$), the Weyl scalar $\Psi=-M/r^3$ is real, and all functions $f_K$ become functions of $r$ alone. The integrability conditions \eqref{compatcond} reduce to $0=0$ and are trivially satisfied. This is a necessary (though not sufficient) condition for the existence of a deformed Carter constant. 

The PDE system \eqref{PDEsystem}, according to these properties, boils down to $\partial_i A_K = 0$ for all $K = 0, 1, 2$ and $i= r, \theta$: all functions $A_K$ in \eqref{Q1final} must be constants with respect to $(r, \theta)$. Inserting this into \eqref{Q1final} simply means, ultimately, that $\Qgeo$ remains a constant of motion for arbitrary tidal couplings $(\cE,\cB)$. This is consistent with the Carter constant $\Qgeo$ reducing to the particle's total angular momentum, an invariant whose existence is a consequence of the Schwarzschild enhanced SO(3) isometry. Our method thus consistently recovers this classical result, and the tidally-perturbed dynamics remains integrable in Schwarzschild, with the four independent constants of motion $(\muT,E,L,Q_0)$.

%%%%%%%%%%%%%%%%%%%%%%%%%%%%%%%%%%%%%%%%%%
\section{Numerical evidence for chaos} \label{sec:chaos}
%%%%%%%%%%%%%%%%%%%%%%%%%%%%%%%%%%%%%%%%%%

The analytical results of Sec.~\ref{sec:integrability} establish the non-existence of a polynomial-in-momenta deformed Carter constant for the tidal Hamiltonian in Kerr. In this section we provide independent confirmation by exploring the phase space directly, through three complementary numerical diagnostics:
\begin{enumerate}[leftmargin=*, align=left, labelindent=0pt, itemindent=0pt, labelsep=0.5em]
    \item Poincar\'e sections, offering a \emph{global and geometric} view on phase space structures (Figs.~\ref{fig:turningpoints}--\ref{fig:zoom});
    \item Lyapunov exponents, providing a \emph{local and temporal} measure of the rate at which neighboring phase space orbits diverge (Fig.~\ref{fig:lyapunov});
    \item escape-time maps, recording whether (and when) orbits plunge into the black hole and illustrate the fractal dependence on initial conditions (Fig.~\ref{fig:escape}).
\end{enumerate}
Importantly, these three diagnostics are not redundant: a positive Lyapunov exponent exclusively establishes sensitivity to initial conditions, a Poincar\'e section displays phase space global structures, and the escape-time map touches on a physical question, namely, does the body plunge, and when? 

Our presentation is deliberately pedagogical, and also aims at revisiting classical numerical phase space dignostics that can be found in the literature. We first lay out the geometrical and numerical setup (Sec.~\ref{ssec:setup}); we then revisit Kerr geodesics and their turning-point structure (Sec.~\ref{ssec:geodesics}): this provides a baseline against which the tidally-perturbed orbits can be discussed. Our results are then presented in the following sections: construction of Poincar\'e sections (Sec.~\ref{ssec:poincare}); computation of Lyapunov exponents (Sec.~\ref{ssec:lyapunov}); and building of the escape-time map (Sec.~\ref{ssec:escape}). 

An additional numerical analysis is discussed in Sec.~\ref{ssec:order}, where we make sure that observed features are of leading order in tidal effects, and not of sub-leading order (which exist but are not under our control) or numerical (which are nonphysical). 

% ============================================================
\subsection{Geometrical and numerical setup} \label{ssec:setup}
% ============================================================

All analyses and resulting plots in this section share the parameters $(a/M, E/\muT, L_z/(M\muT)) = (0.98,\, 0.95,\, 2.1)$. Boyer-Lindquist coordinates $(t,r,\theta,\phi)$ are used throughout. The Kerr mass is set to $M=1$, fixing the units of any proper time and coordinate radius values. The conserved mass $\muT$, which factorizes from all equations of motion, is also set to 1 without loss of generality. The relatively high value of the Kerr spin $a$ is chosen to highlight chaotic effects, which arise when orbits visit the close neighborhood of the black hole horizon region, located at (coordinate) radius $r:=M+\sqrt{M^2-a^2} \simeq 1.2$. 

Note, however, that chaotic artifacts exhibited in this section concern \emph{elliptic} orbits whose periapses reach the deep-field region $r\gtrsim 1.2$, but also go out of the potential well rather far, with typical apoapses located around $r \simeq 18$. Thus, we can say that chaotic orbits/effects are not so much \emph{located} in the deep-field region than they are \emph{seeded} by it: chaotic orbits can still go way out of the central region, as long as they also visit deep in it.

Although the complete phase space is 8D, covered by $(t,p_t,r,p_r,\theta,p_\theta,\phi,p_\phi)$, any orbit (i.e., solution to Hamilton's equations \eqref{HamEq}) is confined to lower-dimensional sub-manifolds where $E=-p_t$ and $L_z=p_\phi$ are fixed. We thus work on the reduced 4D phase space covered by coordinates $(r,p_r,\theta,p_\theta)$. To explore it, the equations of motion~\eqref{tidalODES} for the four variables $(r,p_r,\theta,p_\theta)$ are integrated using an explicit 8th-order Runge--Kutta scheme. In practice, we parametrize the strength of the tidal perturbation by a single bookkeeping parameter $\epsilon$, writing the tidal Hamiltonian as
\begin{equation}
\Htidal = \mu \bigl( c_\text{T}\, \mcE^2 + c_\text{H}\, \text{Re}[\Psi^2] \bigr),
\end{equation}
a form equivalent to \eqref{Htidal} upon using $\mcB^2=\mcE^2+6\,\text{Re}[\Psi^2]$, cf.\ \eqref{Kretschmann}, with the dictionary $c_\text{T}=-(\cE+4\cB)/12$ and $c_\text{H}=-2\cB$. We selected arbitrary values for these coefficients to produce our analyses, namely 
\begin{equation} 
(c_\text{T},c_\text{H}) = (3\epsilon,-2\epsilon),
\end{equation}
for some small number $\epsilon$. We found that $\epsilon=1.5\cdot10^{-2}$ led to effects that were easy enough to locate, measure and interpret. This choice corresponds to tidal parameters $(\cE,\cB)=(-0.6,0.015)$, but we stress that these are chosen for illustrative purposes: our goal is to provide numerical confirmation that chaotic effects arise in phase space, confirming the lack of fourth constant of motion. Repeated analysis with different values of $\epsilon$ produced similar features with the expected scaling $O(\varepsilon)$. We refer to the discussion in \ref{ssec:order} for a quantitative discussion on this particular matter. 

\begin{figure}[h]
\includegraphics[width=.95\linewidth]{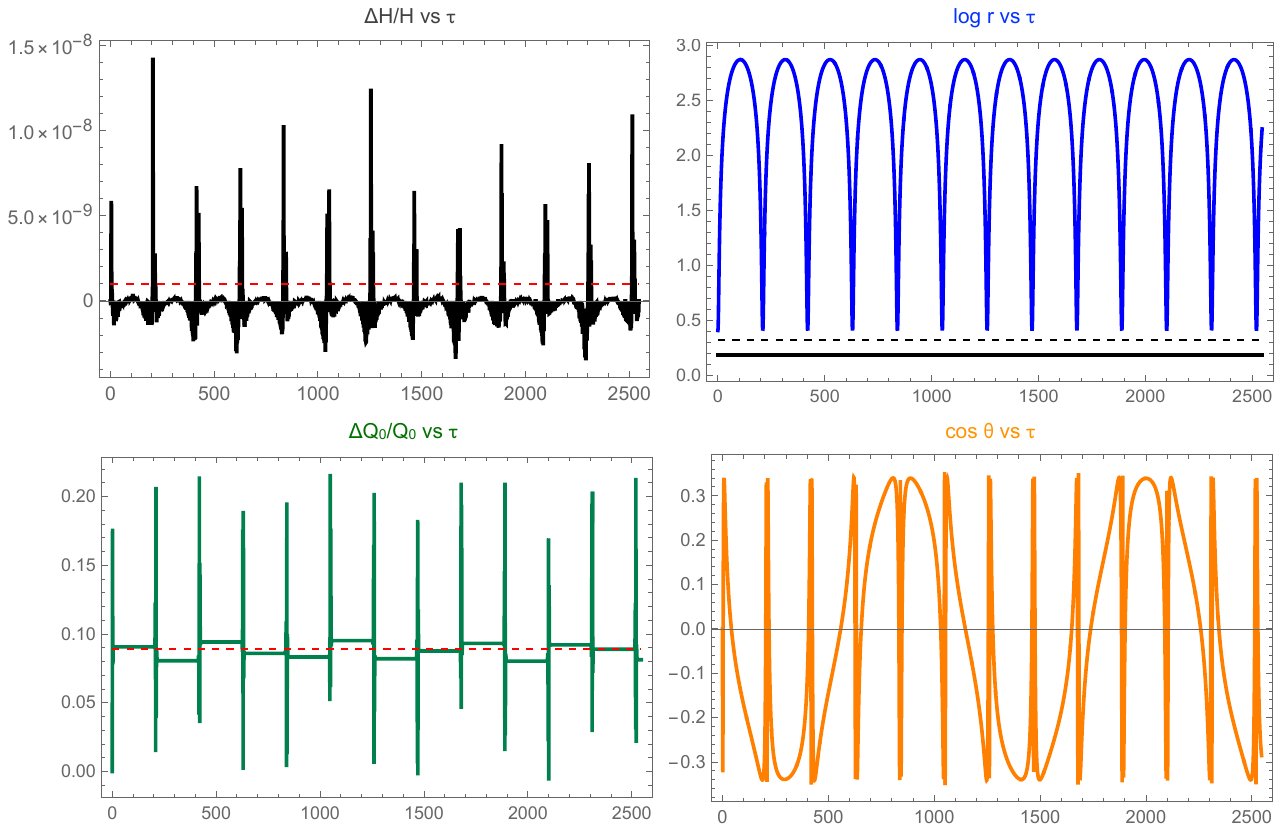}
\caption{Numerical accuracy and trajectory for a tidally-perturbed orbit in Kerr spacetime. Initial position $(r_0,\theta_0) = (1.5,\pi/2)$, with $p_r=0$ and $p_\theta$ chosen such that $\muT=1\Leftrightarrow H=-1/2$. The integration spans a proper time $\tau \in [0,\, 2546]$. \textit{Left column:} relative drift $\Delta H/|H|$ of the total Hamiltonian (black); and relative drift $\Delta \Qgeo/|\Qgeo|$ of the geodesic Carter constant (green). \textit{Right column:} $\log r(\tau)$, with the solid line marking the Kerr horizon and the dashed line the tidally-perturbed unstable spherical orbit radius; (iv) $\cos\theta(\tau)$. Dashed red lines in panels~(i)--(ii) indicate rms values, notice the several orders of magnitude between the true constant of motion $H$, and the non-conserved $Q_0$.}
\label{fig:diagnostics}
\end{figure}

Throughout the analyses, the total Hamiltonian $H$ serves as a diagnostic: its conservation along trajectories is monitored throughout, with typical relative drifts $|\Delta H|/|H| \lesssim 10^{-10}$ for both the geodesic and the tidally-perturbed cases. As a comparison, the geodesic Carter constant is conserved with $|\Delta \Qgeo|/|\Qgeo| \lesssim 10^{-10}$ in the geodesic case, but drifts at the level $|\Delta \Qgeo|/|\Qgeo| \simeq 10^{-1}$ once the tidal perturbation is included, as is expected. A typical tidally-perturbed orbit is depicted in Fig.~\ref{fig:diagnostics}, along with several diagnostic plots reflecting those orders of magnitudes.

% ============================================================
\subsection{Kerr geodesics} \label{ssec:geodesics}
% ============================================================

Before turning on the tidal perturbation $\Htidal$, let us first discuss some properties of Kerr geodesics, in order to anticipate the phase space structures of the full, perturbed system.

Bound Kerr geodesics in the reduced phase 4D space $(r,p_r,\theta,p_\theta)$ are characterised by two constants of motion: the Carter constant $\Qgeo$ and the Hamiltonian $\Hgeo = -\mu^2/2$. These render the 2D system Liouville-integrable, cf. Sec.~\ref{ssec:geoham}. There are two (intertwined) consequences of this: quasi-periodicity of the motion and separability of the ODEs. 

\subsubsection{Quasi-periodic motion} 

\begin{figure}[h]
\includegraphics[width=.95\linewidth]{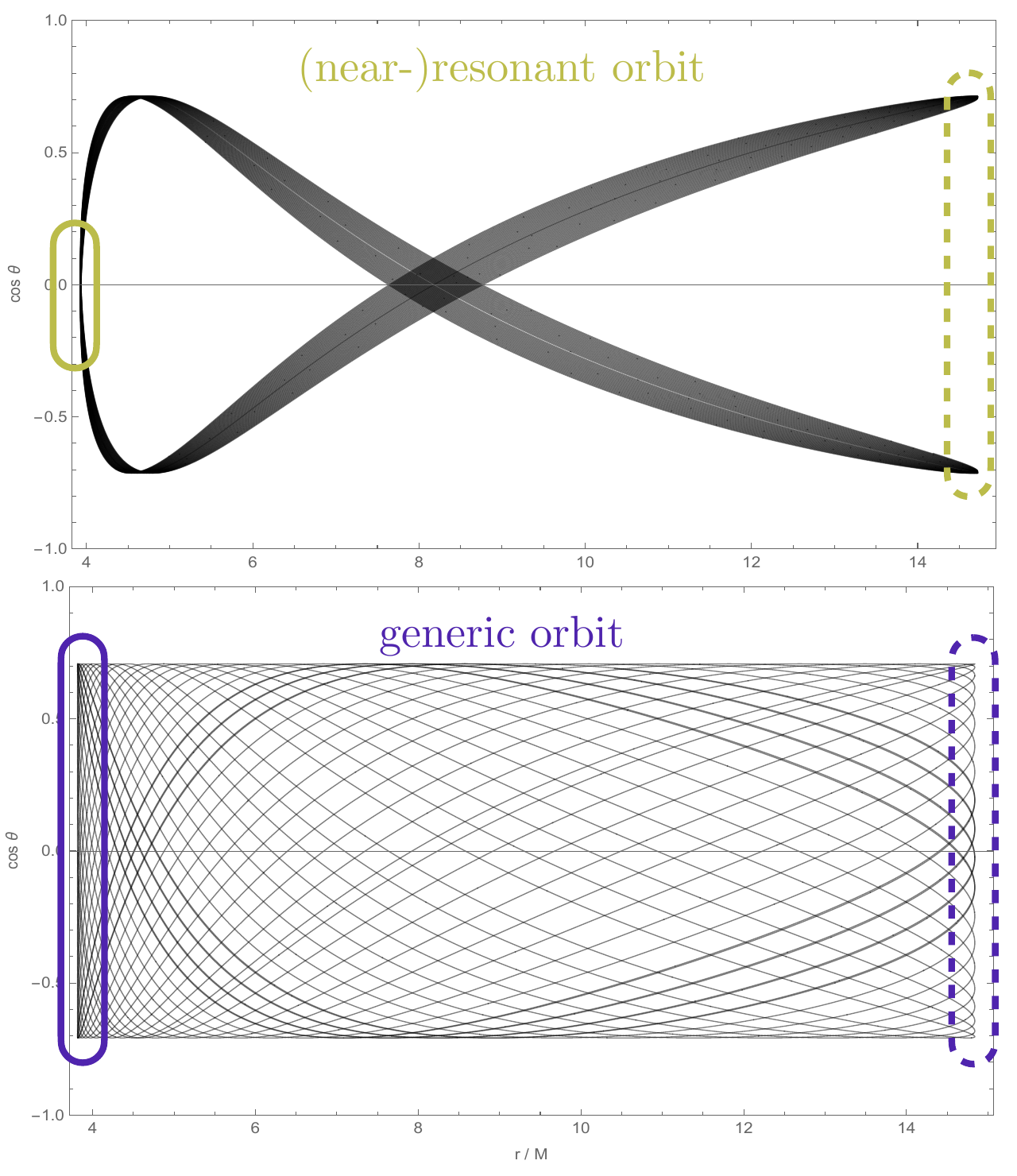}
\caption{Two bound Kerr geodesics in the $(r,\cos\theta)$ configuration space, sharing the same parameters except the initial radius. \emph{Top:} a near-resonant orbit with $\Omega_r/\Omega_\theta \simeq 3/4$ ($r_0 \simeq 3.93$), whose trajectory nearly closes after two radial and three polar oscillations; periapsis and apoapsis passages are marked by solid and dashed circles, respectively. \emph{Bottom:} a generic orbit ($r_0 = 3.814$) with $\Omega_r/\Omega_\theta \notin \mathbb{Q}$, densely filling the box $[r_p, r_a] \times [-\cos\theta_\mathrm{min}, \cos\theta_\mathrm{min}]$. The turning points of both two orbits are part of the 15 orbits depicted in Fig.~\ref{fig:turningpoints}, with matching color (purple and olive).}
\label{fig:orbits}
\end{figure}

The quasi-periodicity of Kerr geodesics means that each bound solution to the geodesic equation traces a trajectory on an invariant 2-torus in the reduced 4D phase space, characterized by two \emph{fundamental frequencies} of motion ,$\Omega_r$ and $\Omega_\theta$ \cite{Sc.02}. When depicted in the 2D configuration space $(r,\cos\theta)$, the orbit either fills a 1D curve (when it closes on itself, i.e., when the frequency ratio is rational, $\Omega_r/\Omega_\theta = p/q \in \mathbb{Q}$ or it densely fills a 2D box $r \in [r_p, r_a]$, $\cos\theta \in [-\cos\theta_\mathrm{min}, \cos\theta_\mathrm{min}]$, in the generic case $\Omega_r/\Omega_\theta \notin \mathbb{Q}$. The former are called \emph{resonant} orbits, and the latter \emph{generic} orbits. An example of each kind is depicted in Fig.~\ref{fig:orbits}.

\subsubsection{Constant turning points} 

The other feature of Kerr geodesics, central to our analysis, is the \emph{separability} of the equations of motion: thanks to the two constants of motion $H_0$ and $Q_0$, the radial and polar equations can be decoupled into two independent first-order ODEs \cite{Schm.02,Mino:2003yg}:
\begin{subequations}\label{GeoODEs}
    \begin{align}
    \dot{r}^2 &= \left(a L_z-E \left(a^2+r^2\right)\right)^2+\Delta(r) \left(\Qgeo-\mu ^2 r^2\right) , \label{Rpotential}\\
    \dot{\theta}^2&= a^2 \left(E^2-\mu ^2\right) \cos^2\theta -L_z^2\cot^2\theta - \mathcal{K}_0 , \label{Thetapotential}
    \end{align}
\end{subequations}
where $\mathcal{K}_0:=\Qgeo + (L_z-a E)^2$ is a shifted Carter constant, $\Delta(r)=r^2-2Mr+a^2$, and the overdot (in this section only) denotes differentiation with respect to the Mino time $s$, defined by\footnote{The equations of motion in Kerr are not separable in proper time, but they are in Mino time \cite{Mino:2003yg}.} $\ud\tau=\Sigma(r,\theta)\,\ud s$. Because they are decoupled, the radial turning points (periapsis $r_p$ and apoapsis $r_a>r_p$) and the polar turning points $\pm\cos\theta_\mathrm{min}$, both defined by the vanishing of the right-hand sides of \eqref{GeoODEs}, are \emph{constants of motion}: they are invariant along the orbit once given set of initial conditions, and can be expressed solely in terms of $(E, L_z, \mu, \Qgeo)$.

\begin{figure*}[ht!]
\includegraphics[width=\linewidth]{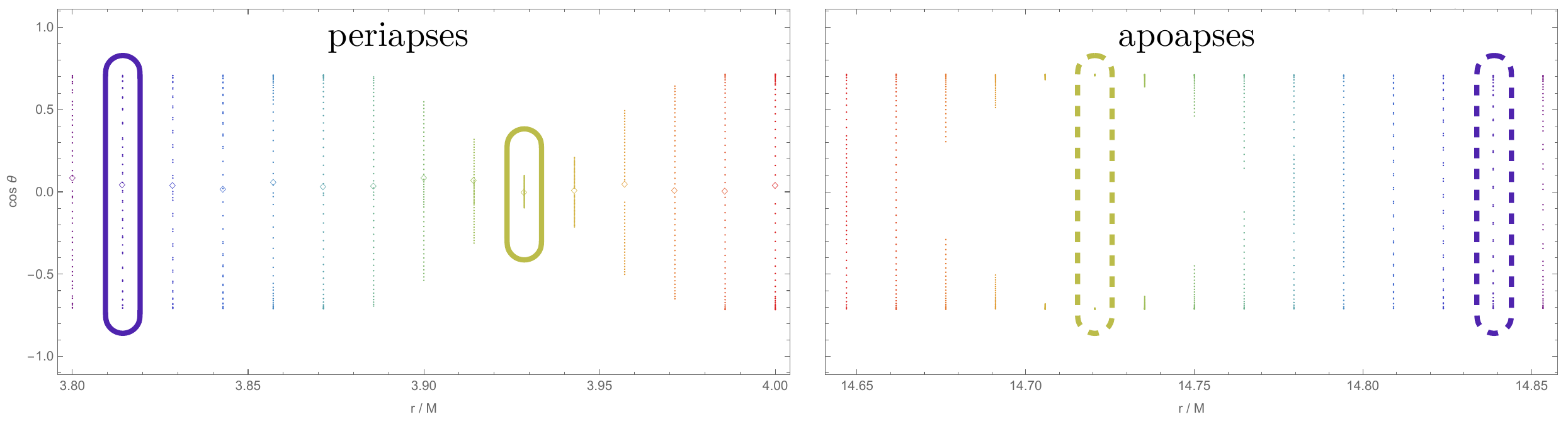}
\caption{Radial turning points of fifteen bound Kerr geodesics differing in initial radius $r_0 \in [3.8,4]$ and random polar angles (diamonds). Orbits are integrated over $\tau\in[0,10^4]$ and recorded in the $(r, \cos\theta)$ plane. Points of the same color belong to the same orbit. Each orbit contributes two vertical segments [one at $r_p$ (periapses, left) and one at $r_a$ (apoapses, right)] confirming that radial turning points are constants of motion, a direct consequence of the separability \eqref{GeoODEs}. Turning points of the two orbits depicted in Fig.~\ref{fig:orbits} are circled, matching the colors (purple and olive).}
\label{fig:turningpoints}
\end{figure*}

The constancy of all turning points has a direct visual signature, which we exploit throughout this section. If, instead of plotting the full orbit, one records only the points $(r,\cos\theta)$ at which the orbit reaches a radial turning point ($p_r\propto \dot{r}=0$), each geodesic produces exactly two vertical segments in the $(r, \cos\theta)$ plane: one at each turning radius, extending over the full range of $\cos\theta$ visited by the orbit. This is illustrated in Fig.~\ref{fig:turningpoints} for fifteen geodesics sharing the same $(E, L_z,\mu)$ but differing in initial radii: each orbit contributes a pair of vertical segments, and the plane is foliated by such pairs. For the near-resonant orbit of Fig.~\ref{fig:orbits}, turning points cluster at a near-discrete set of points instead. One can read off the plot in Fig.~\ref{fig:orbits} that $(r_p,r_a)\simeq(3.93,14.72)$. This corresponds to an eccentricity and semi-latus rectum $$e:=\frac{r_a-r_p}{r_a+r_p},\quand p:=\frac{2r_ar_p}{r_a+r_p}$$ of $(e,p)\simeq (0.58,6.20)$. In turn, these parameters lead to frequencies of $(\Omega_r,\Omega_\theta)\simeq(2.59,3.45)\cdot 10^{-2}$, whose ratio is $\Omega_r/\Omega_\theta\simeq 0.75$, as expected for a $3/4$-resonant orbit. The trajectories and frequency values are generated with our own numerical integrator, and have been checked against those produced by the \texttt{KerrGeodesics} package of the Black Hole Perturbation Toolkit \cite{BHPToolkit}.

\subsubsection{Spherical orbits}

Spherical orbits are a special class of Kerr bound orbits for which the radial coordinate remains constant, $r(\tau) = r_s$, while the polar angle $\theta$ still oscillates between $\pm\theta_\mathrm{min}$~\cite{StWa.20}. In phase space, a spherical orbit corresponds to a fixed point of the radial dynamics, i.e., a double root of the polynomial on the RHS of Eq.~\eqref{Rpotential}. For given $(E, L_z)$, there can either be 0, 1 or 2 spherical orbits. In the case when there are two, there is a stable spherical orbit (SSO), for which small radial perturbations remain bounded, and an unstable spherical orbit (USOs), for which they grow. The USO at radius $r_\mathrm{USO}$ acts as a \emph{separatrix} in phase space: generic bound orbits can only exist for $r_p > r_\mathrm{USO}$. In the Poincar\'e sections below, the USO therefore sets the inner boundary of the region accessible to bound orbits, and its neighbourhood is where the most significant chaotic activity develops, as discussed in Sec.~\ref{ssec:poincare}.

% ============================================================
\subsection{Poincar\'e sections} \label{ssec:poincare}
% ============================================================

We now turn on the tidal effects. A generic, i.e., tidally-perturbed, bound orbit in Kerr corresponds to a continuous 1D curve in the 4D reduced phase space $(r,p_r,\theta,p_\theta)$. Poincar\'e sections are a means to visualize these curves, by projecting from the 4D space down to a 2D one. This is done in two steps.

The first reduction leverages the fact that every trajectory lies on the 3D constant-Hamiltonian hypersurface 
\begin{equation}\label{H=muT}
    H(r,\theta,p_r,p_\theta) = -\muT^2/2,
\end{equation}
which fixes one combination of the phase-space variables; equivalently, one can solve for $p_\theta$ in terms of the other three coordinates along a trajectory. The corresponding 3D-hypersurface is the box depicted on the left of Fig.~\ref{fig:twopoincare}, in grey. In practice, we use $\muT=1\Leftrightarrow H=-1/2$ since the conserved mass factors out of all equations once they are put in normalized form. 

The second reduction defines the Poincar\'e section itself. We select a codimension-one surface $\Pi$ within the aforementioned 3D-hypersurface, by imposing an additional condition on one of the phase-space coordinates. We then record the values of the remaining two phase space coordinates each time the orbit pierces $\Pi$ transversally. The resulting sequence of points on the 2D-surface $\Pi$ defines the \emph{Poincar\'e return map}, whose qualitative structure encodes properties of the full orbits in 4D-space \cite{Arn,ZeLuWi.20}. Of course, different sections $\Pi$ are possible.  

\subsubsection{Choice of Poincar\'e section}

The relativistic chaos literature has predominantly adopted the equatorial plane $\theta = \pi/2$ as a Poincaré section, recording the radial coordinate and momentum $(r, p_r)$ at each crossing. These studies include (but are not limited to) the following references \cite{Bomb.al.92,Suzuki:1996gm,ApoLukCon.09,LukApoCon.10,ConLukApo.11,DesSuvKok.21,DesKok.21,DesKok.23,SuzMae.97,Han.08,ZelLukWitKop.20,BroCarSte.23,TakKoy.09,ShaKolZelLuk.26}. This convention comes from the standard choice in the classical mechanics literature, where the section is typically a coordinate plane in configuration space. 

In our analysis, we chose instead another Poincar\'e section. In essence, it is given by the condition 
\begin{equation}\label{pr=0}
    p_r = 0,
\end{equation} 
and thus records the \emph{spacetime} positions $(r,\cos\theta)$ every time the orbit is at a \emph{turning point}, since $p_r\propto \dot{r}$.\footnote{This is not true for generic Hamiltonians, and not immediate for ours. However, one can show that both brackets in the RHS of \eqref{tidalODES} are $\propto \mathcal{Z}_{\beta\gamma}\hp^\beta h^{\gamma\alpha}$ [recall \eqref{weirdID}]. Setting $\alpha=r$ gives $\propto  \mathcal{Z}_{\beta}^{\phantom{\beta} r}\hp^\beta + \mathcal{Z}_{\beta\gamma}\hp^\beta \hp^\gamma \hp^r$. The second is $\propto p^r$, and so is the first because $\mathcal{Z}_{\beta}^{\phantom{\beta} r}=\mathcal{Z}_{r}^{\phantom{r} r}\delta_\beta^r$, owing to \eqref{decompK} and the diagonality of $g_{ab},K_{ab}$ in the $(r,\theta)$ sector, cf. Sec.~\ref{sssec:basis}. \label{fn:rdot}} In more details, our Poincar\'e section is defined as the 2D-surface
\begin{equation} \label{defPi}
    \Pi := \left\{ (r,\theta,p_r,p_\theta) \in \RR^4, \text{ s.t. } \eqref{H=muT} \text{ and } \eqref{pr=0} \text{ hold} \right\} 
\end{equation}
of the 4D phase space spanned by $(r,\theta,p_r,p_\theta)$. Two connected components of $\Pi$, distinguished by the sign of $\dot{p}_r$, correspond respectively to periapses ($\dot{p}_r > 0$) and apoapses ($\dot{p}_r < 0$);\footnote{These are our definitions of periapsis and apoapsis. They are local in time, and may not successively coincide in general, unlike for geodesics where they are true constants of motion.} they appear as separate point clouds, with periapses always lying on the left, and apoapses on the right, given that $r$ increases from left to right on all figures.

\begin{figure*}[t]
\includegraphics[width=.8\textwidth]{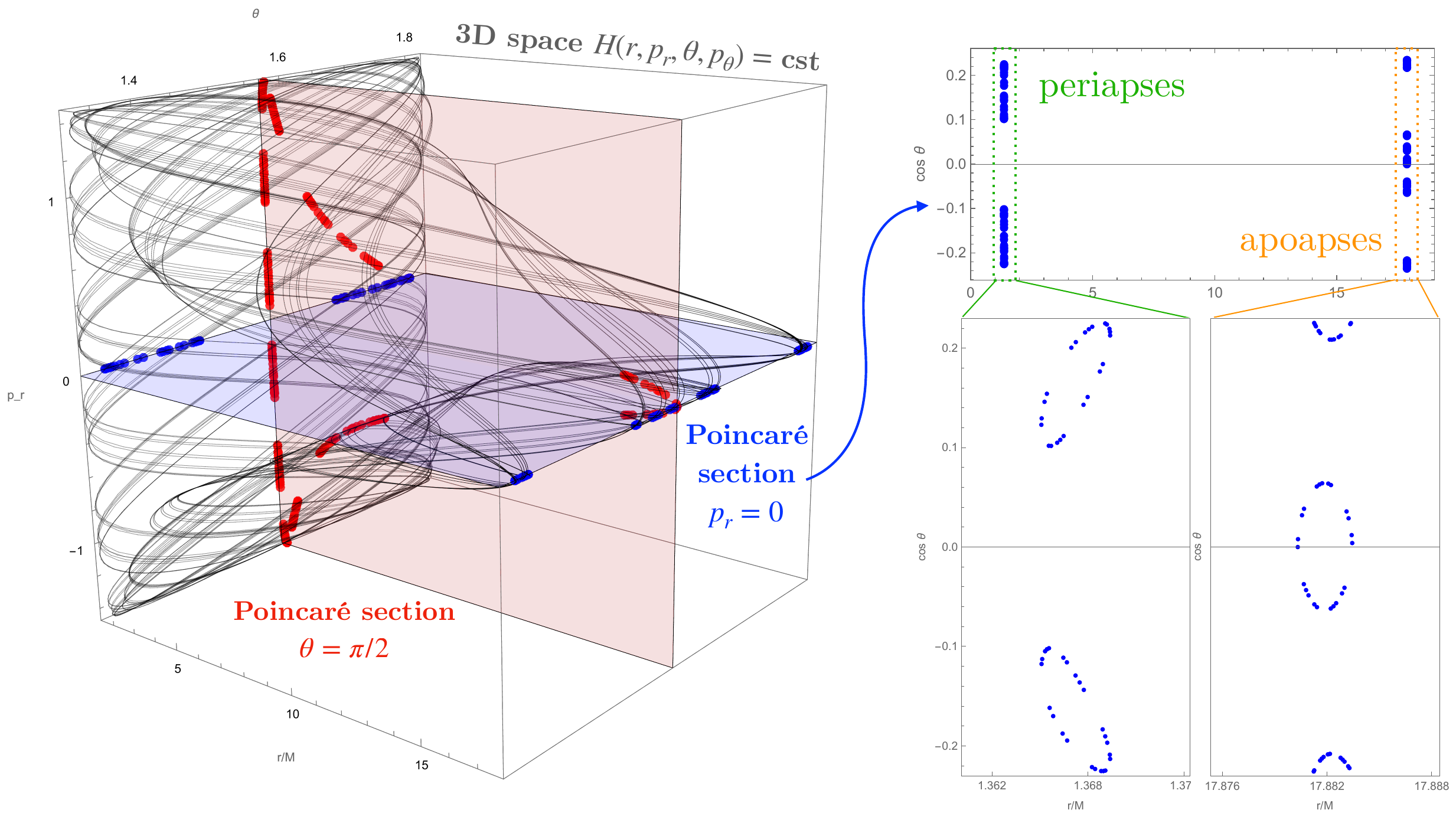}
\caption{Phase-space representation of a bound orbit of the tidally-perturbed system with initial radius $r_0 = 1.5$. \textit{Left:} the orbit as a continuous curve on the energy surface $H(r,p_r,\theta,p_\theta)=\text{cst}$. The blue (resp.\ red) shaded plane indicates the section surface $\Pi:p_r = 0$ (resp.\ $\theta = \pi/2$). \textit{Right:} points $(r, \cos\theta)$ recorded each time the orbit pierces our chosen Poincar\'e section transversally, i.e.\ at radial turning points ($p_r = 0$). The points \emph{appear} to lie on vertical lines, just for the geodesic case depicted in \ref{fig:turningpoints}. However, a zoom reveals that successive periapses and apoapses are not distributed vertically, and rather fill a finite radius interval, i.e., a loop in $(r,\cos\theta)$-space.}
\label{fig:twopoincare}
\end{figure*}

Two considerations motivated our choice of section:

\paragraph*{(i) Configuration-space interpretation.}
On the section $\Pi: p_r = 0$, the recorded coordinates $(r,\cos\theta)$ are the literal spacetime positions at radial turning points of the orbit, not abstract phase-space slices. One can directly infer the radial excursion of the orbit in the $(r,\cos\theta)$ plane, and relate features of the section to the spatial geometry of the trajectory.

\paragraph*{(ii) Direct visual signature of separability.}
For Kerr geodesics, the existence of the Carter constant renders the radial and polar motions separable: the turning-point radii $r_p$ and $r_a$ are independent of $\theta$. On $\Pi$, each geodesic therefore produces a pair of strictly vertical segments: this is precisely the content of Fig.~\ref{fig:turningpoints}, which is nothing but our Poincar\'e section of the geodesic flow on $\Pi$. These vertical segments \emph{are} the level sets of the Carter constant $\Qgeo$: a geodesic with a different value of $\Qgeo$ (at the same $E$, $L_z$) traces a different pair of vertical segments at different radii, and the entire section is foliated by such pairs. Their deformation or dissolution into scattered points under a perturbation provides a direct visual signature of loss of integrability.

Our section $\Pi$ defined in \eqref{defPi}, and the comparison with the more classical $\theta=\pi/2$ section, are illustrated on the left in Fig.~\ref{fig:twopoincare}, where they correspond to the blue and red slices, respectively. Points belonging to the sections are highlighted in the same color. 

\subsubsection{Reading of a Poincar\'e section}\label{sec:configs}

The qualitative character of an orbit can be read off directly from its trace on $\Pi$. There are three classes:
\begin{enumerate}[leftmargin=*, align=left, labelindent=0pt, itemindent=0pt, labelsep=0.5em]
    \item \emph{1D smooth curve} (generic orbit): the orbit lies on an invariant 2-torus; its intersection with $\Pi$ is a continuous closed curve. An example is the set of points\footnote{In practice the continuous nature is obtained after an infinite amount of crossings. Our plots show instead a dense filling of points forming a curve asymptotically.} in purple in Fig.~\ref{fig:turningpoints};
    \item \emph{Finite set of isolated points} (resonant orbit): the orbit closes exactly on itself after a finite number of radial and polar librations. Therefore, a finite number of points on $\Pi$ are repeatedly visited during the orbital evolution. An example is the set of points in olive in Fig.~\ref{fig:turningpoints};
    \item \emph{2D scattered cloud} (chaotic orbit): the orbit's trace fills a 2D region of $\Pi$, with no discernible curve structure. This is reminiscent of chaos, and it does not happen for Kerr geodesics, which are integrable.
\end{enumerate}

The connection between the Poincar\'e section and integrability can be made more precise via the Liouville--Arnold theorem \cite{Arn}. Suppose that a smooth function $C(r,\theta,p_r,p_\theta)$, independent of $H$ and in involution with it ($\{C,H\}=0$), existed on an open region $\mathcal{U}$ of the reduced phase space. Then on $\mathcal{U}$, the joint level sets of $H$ and $C$ would be compact 2D-surfaces diffeomorphic to 2-tori, and every orbit would be confined to one such torus. The intersection of each 2-torus with the section $\Pi$ is generically a 1D closed curve. In particular, $\Pi\cap\mathcal{U}$ would be \emph{foliated} by continuous curves, with no orbit filling a 2D region.

Contrapositively, the observation of a densely filled 2D region on $\Pi$ rules out the existence of a smooth second integral on that region of phase space. This is stronger than the analytical result of Sec.~\ref{sec:integrability}, which excludes integrals of the polynomial-in-momenta class (albeit on the \emph{entire} phase space, not just on a particular region). The two approaches are complementary: the analytical proof is global but restricted to a specific functional class; the numerical evidence applies to all smooth integrals but is confined to the specific orbits and parameter values explored.

Regardless, having introduced all necessary tools (numerical implementation in \ref{ssec:setup}, illustration of the geodesic phase space \ref{ssec:geodesics} and the Poincaré section \ref{ssec:poincare}), we can now exhibit and discuss our findings. 

% ============================================================
\subsection{Results: Poincaré section} \label{ssec:resultpoincare}
% ============================================================

\begin{figure*}[ht]
\includegraphics[width=\textwidth]{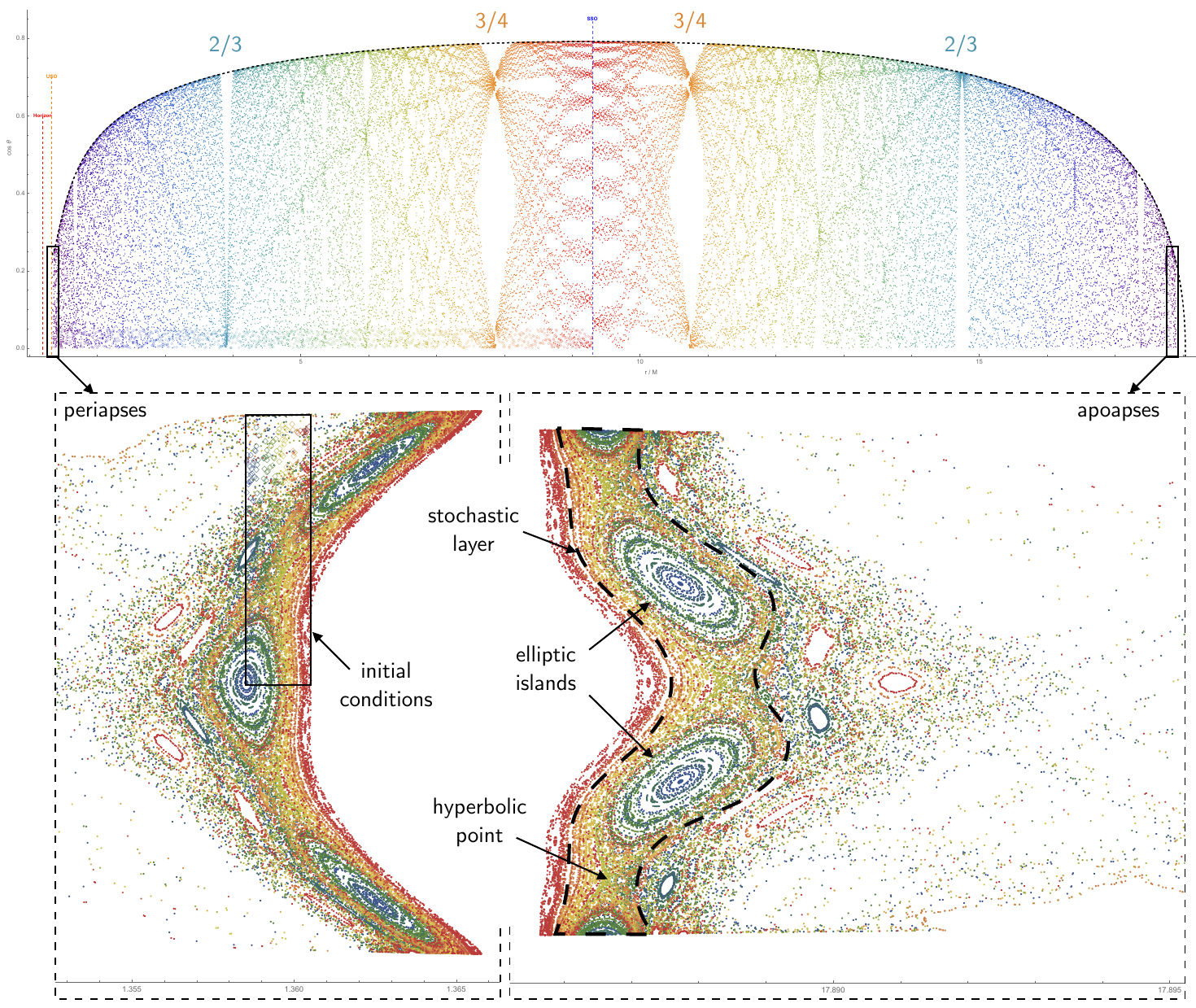}
  \caption{\emph{Top:} Poincar\'e section $\Pi:p_r=0$ for the tidally-perturbed Kerr dynamics (only the top-half $\cos\theta\geq0$ is shown). One can see the location of the most resonant orbits ($2/3$ and $3/4$) as well as higher-order resonances (note that stronger, lower-order resonances like $1/2$ or $1/3$ are not reachable for the values of $E,L$ used for our analysis \cite{BrGeHi.15,BrGeHiPRL.15}). The section looks smooth overall with little difference from its geodesic counterpart. \emph{Bottom:} zooms in the inner and outer region of allowed bound orbits, depicting the successive periapses (left) and apoapses (right) of a sample of orbits with initial conditions chosen randomly in the black rectangle. This picture is the tidally-perturbed version of its geodesic counterpart Fig.~\ref{fig:turningpoints}, with non-constant peri/apoapses as in the lower right of Fig.~\ref{fig:twopoincare}. Indicators of a perturbed, chaotic phase space section appear: stochastic layers, elliptic islands, hyperbolic points, as explained in the text in Sec.~\ref{ssec:resultpoincare}.}
  \label{fig:zoom}
\end{figure*}

The geodesic baseline is Fig.~\ref{fig:turningpoints}: all orbits produce either strictly vertical segments or finite sets of points on $\Pi$ (configurations 1. and 2. described in section \ref{sec:configs}).
When tidal effects are turned on, the structure is qualitatively different, and a typical example of Poincaré section is depicted in Fig.~\ref{fig:zoom}. At the top, the large scale picture seems smooth and geodesic-like, with 999 orbits sampled at various initial radii between the tidally-perturbed USO and SSO, and random initial polar angles. Those orbits produce 46864 turning points that fill \emph{seemingly} vertical lines (only the top half $\cos\theta\geq 0$ is displayed). Artifacts from geodesic resonant orbits are also visible, including the $2/3$ and $3/4$ resonant orbits. However, the tidally perturbed orbits in Fig.~\ref{fig:zoom} have the same property as the orbit depicted in the right of Fig.~\ref{fig:twopoincare}: the successive turning points of each orbit fill a closed but \emph{non-vertical} curve. Turning points are thus not constants: the tidal ODEs are coupled and no fourth constant of motion exists, but most orbits are still regular and non-chaotic. These are the celebrated KAM torii: slightly deformed torii expected from perturbed integrable Hamiltonian system theory \cite{Arn}.

However, when zooming in on orbits that visit the vicinity of the strong-field region, with periapses just outside the tidally-perturbed USO, the emergence of chaos can be visibly seen. The bottom two panels of Fig.~\ref{fig:zoom} show 91741 turning points produced by 400 orbits, with initial conditions sampled in the black rectangular region. The maximum integration proper time is $\tau=30000$, although many orbits end up plunging before reaching it. Periapses are on the left, and apoapses on the right, with one color per orbit. 

The main features are the following. Along the deformed curves, one identifies a chain of \emph{elliptic islands}: nested closed loops, (mostly blue and green) each traced by a single orbit (one color), organized around the stable periodic orbits of a broken resonant torus. Between consecutive islands, the curves pinch at \emph{hyperbolic points}, which are unstable periodic orbits of the same resonance. Emanating from these hyperbolic points, thin \emph{stochastic layers} develop (mostly yellow), in which one notices color-mixing: points of distinct colors interleave, meaning that orbits with drastically different initial conditions now visit the same neighbourhoods of $\Pi$, and are no longer confined to individual invariant curves. Further out from the chain, several smaller closed loops (mostly red and blue) float within these layers: these are higher-order resonances. This hierarchy is characteristic of nearly-integrable Hamiltonian systems.

Moving away from the regular region (towards the tidally-perturbed USO on the periapsis branch (left), and towards the largest apoapsis compatible with the Hamiltonian constraint \eqref{H=muT} on the apoapsis branch (right)), the stochastic layers become wider and eventually overlap and merge into a single \emph{connected chaotic sea}, visible on both panels as a 2D cloud of fully interleaved colored points (it also extends beyond the plotted windows). There, no curve structure can be distinguished: a single orbit can typically explore this 2D-region of $\Pi$, and orbits that started arbitrarily close to one another end up spreading over the entire sea. Since one color corresponds to one orbit, the presence of all colors constitutes a direct evidence that the invariant tori have been destroyed in this region.\footnote{This is not in contradiction with the KAM theorem, which only states that a finite measure of tori survive the perturbation, not precluding the existence of chaotic regions. The KAM theorem is a perturbative result, and the size of the chaotic region grows with the perturbation strength. It is large enough to be clearly visible in Fig.~\ref{fig:zoom}.} 

Note, lastly, that the two panels on the bottom depict the same set of orbits: a chaotic orbit possesses both deep periapses and distant apoapses, so that the chaotic sea occupies the left edge of the periapsis branch \emph{and} the right edge of the apoapsis branch. Many such orbits diffuse across the USO and plunge into the black hole before $\tau = 30000$, which is why the sea is comparatively sparsely populated.

%%============================================================
\subsection{Results: Lyapunov exponents} \label{ssec:lyapunov}
% ===========================================================

\begin{figure*}[ht!]
\includegraphics[width=.8\textwidth]{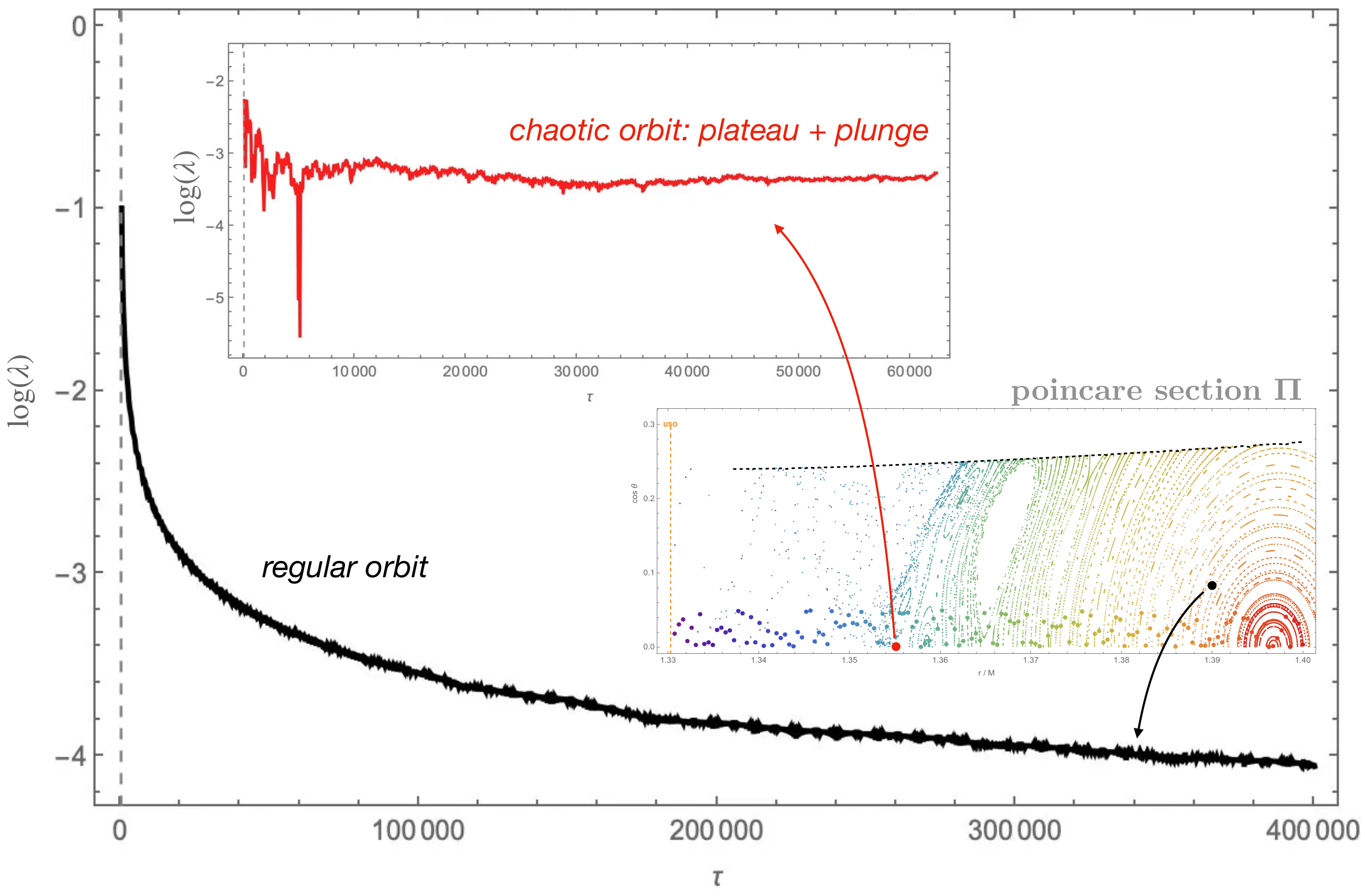}
  \caption{Convergence of the maximal Lyapunov exponent $\lambda_{\max}(\tau)$, plotted as $\log_{10}|\lambda_{\max}|$ versus proper time $\tau$, for two tidally-perturbed orbits, obtained using the Benettin algorithm (see text). Black curve: a regular orbit at $r_0 = 1.39$, on a surviving KAM torus; the exponent decays monotonically over $\tau \sim 4\times 10^5$, confirming quasi-periodic motion. Red curve: a chaotic orbit at $r_0 = 1.355$, in the chaotic sea near the separatrix; after an initial transient, $\lambda_{\max}$ plateaus at $\lambda_\infty \simeq 4\times 10^{-4}$. Arrows point from the insert which depicts the two initial conditions in the Poincaré section $\Pi$.}
  \label{fig:lyapunov}
\end{figure*}

Another hallmark of chaos is the sensitivity to initial conditions, and non-integrable hamiltonian systems are also subject to this feature. A particularly useful tool to measure this sensitivity is the so-called maximal Lyapunov exponent $\lambda_{\max}$, which we computed using the Benettin algorithm~\cite{Skokos2010}. Two nearby initial conditions, separated by $\delta_0\sim 10^{-8}$ in the $p_r$ direction,\footnote{The perturbation is applied to $p_r$ rather than to a coordinate: since $\partial H/\partial p_r = 0$ on the section $p_r=0$, a small offset $\delta_0$ violates the mass-shell constraint only at $O(\delta_0^2)$, whereas a coordinate offset would require re-solving for $p_\theta$ to keep the neighboring orbit on the same constraint surface. The choice is inconsequential for $\lambda_{\max}$ itself, since any generic perturbation aligns with the unstable direction after a short transient.} are integrated forward in chunks of proper-time duration $\Delta\tau$. Once the $k$-th chunk of $\Delta\tau$ is over, the phase-space separation $\delta_k = \|x_2^{(k)} - x_1^{(k)}\|$ (Euclidean-norm on the vector $x=(r,\theta,p_r,p_\theta)$) is computed and recorded, and the perturbed orbit is renormalized back to distance $\delta_0$ from the reference orbit, preserving the separation direction. The running average
\begin{equation} \label{lambdamax}
    \lambda_{\max}(\tau) = \frac{1}{\tau}\sum_{k=1}^{N} \log\frac{\delta_k}{\delta_0}\,,
\end{equation}
with $\tau = N\Delta\tau$, then converges to a positive constant for chaotic orbits and decays for regular ones \cite{ContopoulosBook.02,Skokos2010}.

Figure~\ref{fig:lyapunov} shows the result of calculating $\lambda_{\max}(\tau)$ via the aforementioned method for two tidally-perturbed orbits. They differ in initial conditions only in the starting radius: $r_0 = 1.390$ and $r_0 = 1.355$, just outside the unstable spherical orbit radius.
The radius $r_0=1.390$ seeds a regular orbit (black curve in Fig.~\ref{fig:lyapunov}), with $\lambda_{\max}$ decaying over the full integration interval with no sign of leveling off. This monotonic decay confirms that nearby orbits diverge at most polynomially, as expected for quasi-periodic motion confined to an invariant 2-torus. On the Poincar\'e section, this orbit traces a smooth closed curve (a surviving KAM torus).

The other initial radius $r_0=1.355$ generates a chaotic orbit (red curve in Fig.~\ref{fig:lyapunov}), and $\lambda_{\max}$ exhibits a qualitatively different behavior. After an initial transient, $\lambda_{\max}(\tau)$ converges to a positive plateau at $\lambda_\infty \simeq 4\times 10^{-4}$. Its inverse, $\sim 2.5\times 10^3$, is the time needed for two initially neighboring orbits to separate by a factor $e$ (the \emph{$e$-folding time}), here about twenty radial periods. At $\tau \simeq 6.2\times 10^4$, the orbit plunges into the black hole: chaotic diffusion has carried the particle across the USO. On the Poincar\'e section, such orbits generate the diffuse points in the chaotic sea described above.

Note that the two orbits depicted in \ref{fig:lyapunov} share the same conserved quantities $(H,E,L_z)$, Kerr spin and tidal couplings, and differ in initial radius by only $\Delta r \simeq 0.035$ near the unstable spherical orbit, as seen on the insert in \ref{fig:lyapunov} showing the location of both seeds and the USO on a Poincaré section. Yet, this is enough to separate quasi-periodic motion from deterministic chaos. Once again, a positive maximal Lyapunov exponent is a defining quantitative signature of chaos, and is incompatible with the existence of a smooth integral of motion that would confine orbits to invariant tori on that region of phase space.

% ============================================================
\subsection{Results: Escape-time map} \label{ssec:escape}
% ============================================================

The Poincar\'e sections and Lyapunov exponents diagnose chaos through the geometry of orbits and their local rate of divergence. We present a third and final diagnostic associated to a direct astrophysical phenomenon: the \emph{escape time} $\tau_{\rm p}$, defined as the proper time at which an orbit plunges (whence the $\rm p$ in $\tau_{\rm p}$) into the black hole.\footnote{This method is a relativistic adaptation of a chaos diagnostic used in classical mechanics, see \cite{ContopoulosBook.02} for details.} Numerically, we integrate the equations of motion and trigger when $r$ first drops below $r_{\rm hor} + 10^{-2}$, or at $\tau_{\rm max}=10^4 $ if the orbit remains bound throughout the integration.\footnote{Our threshold $r_{\rm hor}+10^{-2}$ lies $\sim0.12$ below the USO: orbits reaching it have crossed the separatrix and are clearly plunging.} For an extreme mass-ratio inspiral, whether the secondary plunges, and after how many orbital cycles, is an important piece of information contained in the waveform, and the sensitivity of $\tau_{\rm p}$ to initial conditions can be of interest.

\begin{figure*}[ht!]
\includegraphics[width=\textwidth]{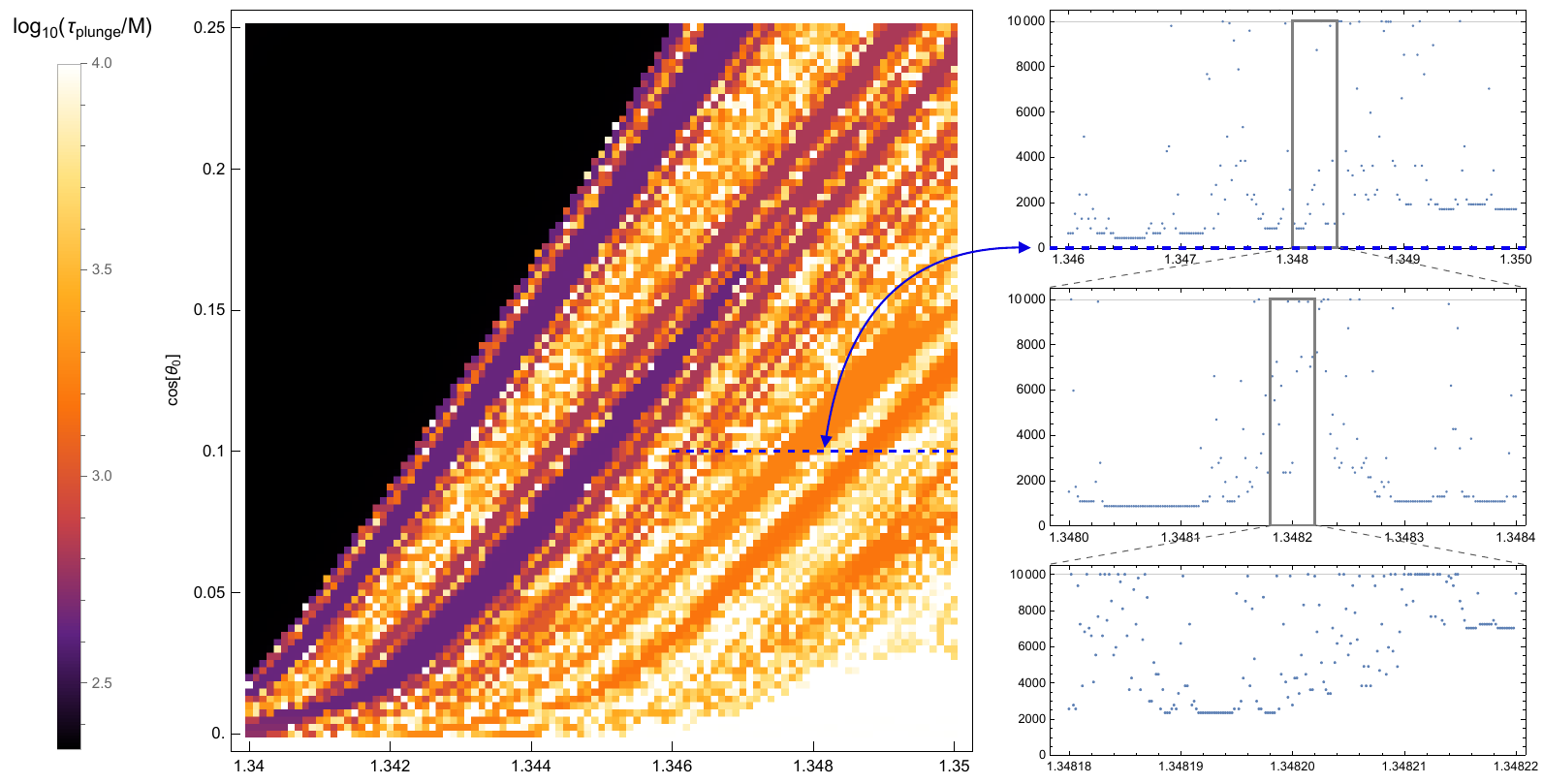}
\caption{\emph{Left:} escape-time map for a selected $101 \times 101$ grid of initial conditions seeding tidally-perturbed Kerr orbits, showing $\log_{10}(\tau_{\rm p})$. The black sector contains only orbits plunging early; the curved bands below the boundary mark successive radial passages on which orbits plunge. The blue dashed line indicates the subset of initial conditions with $\cos\theta_0=0.10$ seeding orbits analysed on the right. \emph{Right, top to bottom:} $\tau_{\rm p}$ for orbits with initial radii sampled at resolutions $\delta r_0 = 2\times10^{-5}$, $2\times10^{-6}$, $2\times10^{-7}$; each gray box marks the sub-interval magnified in the panel below. The escape-time structure recurs at every scale, with a plateau density scaling as $(\delta r_0)^{-1.02}$ (see text).}
\label{fig:escape}
\end{figure*}

Figure~\ref{fig:escape} (left) displays $\tau_{\rm p}$ over a $101\times 101$ grid of initial conditions $(r_0,\cos\theta_0)$ on the section surface $\Pi$, with $r_0\in [1.34,1.35]$ and $\cos\theta_0\in [0,0.25]$. We chose this particular region because two distinct (and expected) features stand out. In the upper-left (black), the region is filled with orbits directly plunging into the black hole within a few radial periods ($T_r \lesssim 200$ with our chosen parameters), and $\tau_{\rm p}$ there varies regularly with the initial data. In the lower-right (colored), below a sharp boundary lies a second region, organized into a family of diagonal stripes. Because plunges can only occur near periapsis, $\tau_{\rm p}$ appears quantized in units of the radial period $T_r$: it takes discrete values, so the function $r_0 \mapsto \tau_{\rm p}$ looks piecewise constant. The stripe edges, however, seem sharp only because of the image resolution: under magnification they dissolve into a more intricate structure. We investigated that structure as follows.

First, to resolve it, we fix $\cos\theta_0 = 0.10$ (the horizontal cut marked in blue in Fig.~\ref{fig:escape}) and scan $\tau_{\rm p}$ with 201 initial radii at three successively finer resolutions, $\delta r_0 = 2\times 10^{-5},\ 2\times 10^{-6},\ 2\times 10^{-7}$, each window (right panels, top to bottom) magnifying the interval boxed in the panel above it. Qualitatively, one can see the alternation between short escape times (plateaus), long-lived orbits (scattered), and survivors $\tau_{\rm p}=\tau_{\rm max}$ (top line). These structures persist at every scale: in particular, refining the resolution does not smooth the structure into clear, definite intervals.

To make this quantitative, we exploit the fact that $\tau_{\rm p}$ is quantized: since an orbit can only cross the plunge threshold near a ``periapsis'' turning point, the escape time takes almost-discrete values, and the map $r_0 \mapsto \tau_{\rm p}(r_0)$ is piecewise constant. Each maximal set of consecutive initial radii sharing a common escape time forms a \emph{plateau},\footnote{Precisely, a plateau is a maximal set of consecutive sampled initial radii over which $\tau_{\rm p}$ is constant, two escape times being counted as equal when they differ by less than $1$, far below the plateau-to-plateau separation of one radial period ($\sim 100$). Isolated points, whose two neighbours have different escape times, count as plateaus of unit length. This is important since near a fractal boundary such singletons proliferate and carry precisely the structure we are measuring.} and we define the plateau density $\rho$ as the number of distinct plateaus per unit $r_0$ within a given scanning window. For a smooth (non-fractal) basin boundary, refining the resolution would eventually resolve the plateaus into a fixed number of wide, well-separated intervals, so that $\rho$ would saturate at a finite value set by the intrinsic scale of the structure. A fractal boundary, by contrast, has no such intrinsic scale: new plateaus appear at every refinement, and $\rho$ grows without bound. Counting the plateaus in each of the three windows above, we find
\begin{equation} \label{eq:plateaudensity}
    \rho \simeq 3.675\times10^{4},\quad 2.975\times10^{5},\quad 3.975\times10^{6}
\end{equation}
for $\delta r_0 = 2\times10^{-5},\ 2\times10^{-6},\ 2\times10^{-7}$ respectively, i.e.\ an increase by roughly one order of magnitude for each decade of refinement. We have thus calculated the number of distinct escape-time plateaus per unit $r_0$: it grows as $\propto (\delta r_0)^{-\gamma}$ with $\gamma \simeq 1.02$ (from a linear fit of $\log\rho$ against $\log\delta r_0$) across the two decades probed. In other words, the plateau width scales linearly with the resolution, with no characteristic scale, which is reminiscent of a fractal set.

We have also made sure that this structure is dynamical, not numerical. The plateau density that we calculated is stable to within $1\%$ under a hundredfold tightening of the integration tolerance, and agrees to within $4\%$ between an explicit eighth-order Runge--Kutta integrator and a stiffness-switching scheme; along bound orbits the Hamiltonian is conserved to $|\Delta H|/|H|\lesssim 10^{-10}$. We can thus confidently say that the escape time exhibits a fractal dependence on initial conditions: plunging and surviving orbits fill this region of phase space, with both outcomes existing at all resolved scales studied.

%%%%%%%%%%%%%%%%%%%%%%%%%%%%%%%%%%%%%%%%%%
\section{Discussion} \label{sec:discussion}
%%%%%%%%%%%%%%%%%%%%%%%%%%%%%%%%%%%%%%%%%%

We conclude this paper with a summary of our results and their place within the broader programme of relativistic integrability (Sec.~\ref{ssec:summary}), a check that the numerically observed chaos is a genuine first-order effect in the tidal coupling (Sec.~\ref{ssec:order}), and a discussion of the astrophysical implications and directions for future work (Sec.~\ref{ssec:prospects}).

\subsection{Summary} \label{ssec:summary}

We have shown that a non-spinning compact object orbiting a Kerr black hole, when endowed with a tidally-induced quadrupole, follows a leading-order motion that is Hamiltonian but \emph{not} integrable. This is the conclusion reached after several sub-results are obtained.

A first result, on which everything else rests, is that this tidal dynamics is Hamiltonian at all. In Sec.~\ref{sec:Ham} we showed that the MPTD equations with a tidally-induced quadrupole \eqref{Jtidal} are generated by the Hamiltonian $H$ of Eq.~\eqref{Htotnew}. The main takeaway is conveniently summarized in Sec.~\ref{ssec:HamSum}. This
holds for \emph{any} background, not just Kerr. Beyond its own interest, this structure is what makes the rest of the analysis possible: only because the dynamics is Hamiltonian can the existence of a conserved Carter constant be phrased as a cohomological equation \eqref{cohomo} and settled through Poisson brackets, rather than by inspection of the equations of motion.

With this in hand, we asked whether the four geodesic constants of motion survive. Three do, in some form: the energy $E$ and axial angular momentum $L_z$, tied to the isometries of the background, and the dynamical mass, which admits a linear-in-tide deformation $\muT$ \eqref{muT} that is exactly conserved at this order. The Carter constant does not: it admits no deformation within the class of polynomial-in-momenta phase-space functions, which is both natural and, as we argued in Sec.~\ref{sec:Carterlike}, exhaustive. The obstruction is geometrical: the integrability conditions \eqref{compatcond} of the cohomological equation \eqref{cohomo}, which must hold for a conserved tidal-corrected Carter constant to exist, are not satisfied for generic couplings $(\cE,\cB)$ and Kerr spin $a\neq0$.

The proof relies on one key intermediate result, established in Sec.~\ref{sec:tidalscalars}: the scalar tidal invariants $\mcE^2$ and $\mcB^2$ can be expressed in closed form \eqref{finalEB} solely in terms of the Weyl scalar $\Psi$ and the normalized geodesic Carter constant $\hat{Q}_0$, their entire momentum dependence being carried by the latter \eqref{E2simp}--\eqref{B2simp}. These formulae, we believe, are new, and could be useful wherever both Kerr orbital dynamics and tidal effects matter. Their derivation rests on a covariant, Killing--Yano-based formulation and extends to the family discussed in \cite{Ra.Iso.Dru.IntegO2.26} of Einstein spaces endowed with a Killing--Yano tensor.

While our proof of the non-existence of a deformed Carter constant is analytic, we also verified the claim numerically, through several diagnostics of the phase-space structure. The goal was twofold. First, to develop numerical tools that will be reused in forthcoming studies. Second, to reveal, as expected in the absence of integrability, the chaotic features of the phase space. We observed the principal signatures of non-integrable dynamics in all tests: Poincar\'e sections, Lyapunov exponents, and escape-time maps.

\subsection{A dynamical characterization of black holes}

Taken together with its spin-induced counterpart \cite{Ra.CQG.24, Ra.Iso.Dru.IntegO2.26}, our result points to a remarkable dynamical characterization of black holes, at least to quadrupole order in the multipole expansion. The spin-induced quadrupole preserves a deformed Carter constant only when its coupling takes the black-hole value $\kappa=1$ \cite{ComDruVin.23,Ra.Iso.Dru.IntegO2.26}; the tidal quadrupole preserves none unless its couplings vanish, $\cE=\cB=0$, which is again the black-hole value. In both channels, then, integrability at quadrupolar order holds precisely for black holes, and is lost for any other body. Whether this is a coincidence of the quadrupolar order or the low-order manifestation of a deeper, all-multipole dynamical uniqueness of black holes is still an open question. The tools assembled here and in \cite{ComDruVin.23,Ra.Iso.IntegO1.26,Ra.Iso.Dru.IntegO2.26} are, in principle, sufficient to settle it order by order; the quadrupolar chapter, at least, is now closed with the present work.

\subsection{On the order of the numerical effects} \label{ssec:order}

\begin{figure}[t]
\centering
\includegraphics[width=\linewidth]{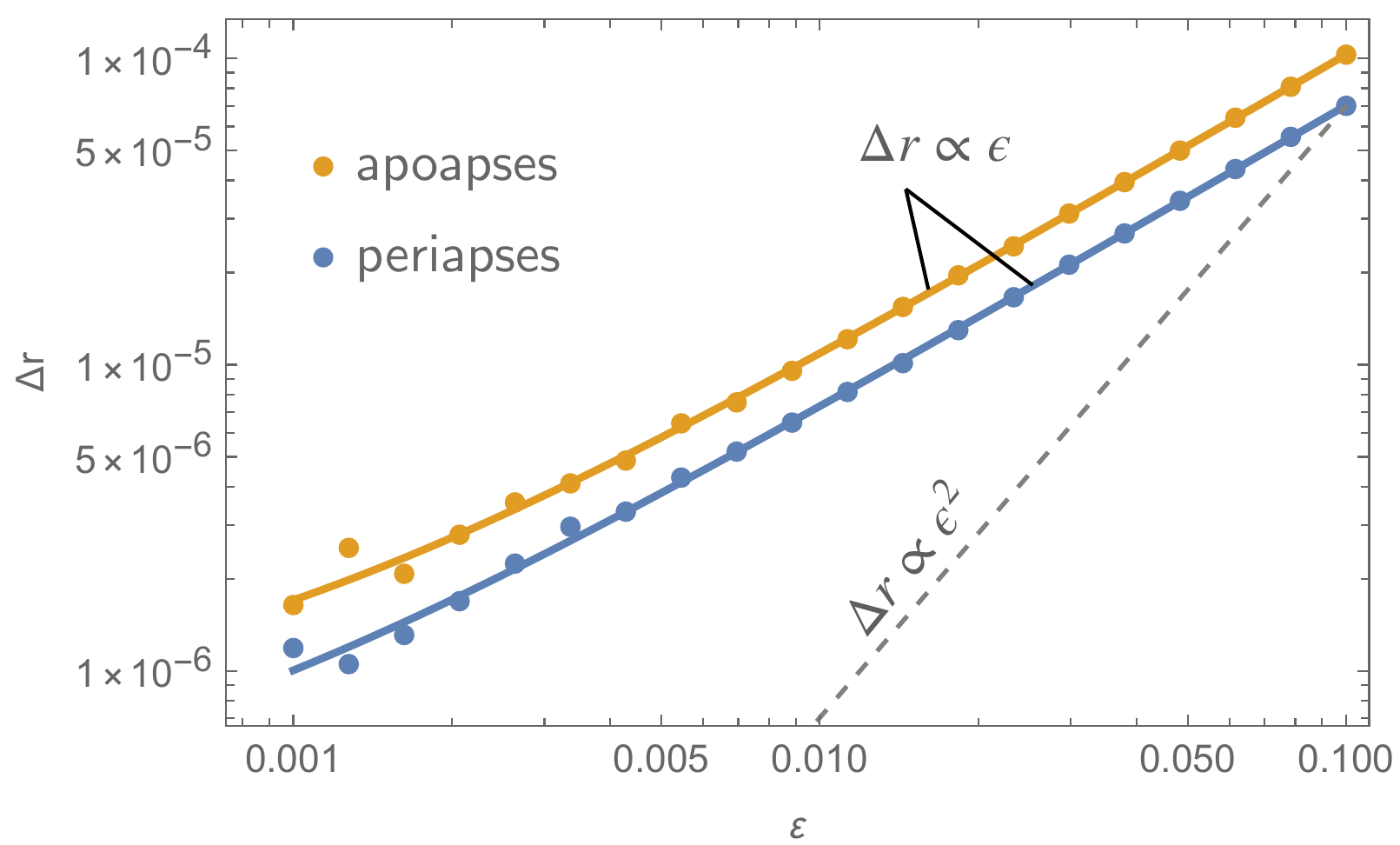}
\caption{\emph{Points:} Radial spread $\Delta r$ of the turning-point locus of a fixed regular orbit ($r_0 = 7$, $\cos\theta_0 = 0.15$) versus the tidal coupling $\epsilon$, for the periapsis ($\Delta r_p$, blue) and apoapsis ($\Delta r_a$, orange) branches, at twenty geometrically-spaced values $\epsilon \in [10^{-3},10^{-1}]$. The integration is over $\tau_{\max}=10^5 $, producing 477 pairs of peri/apoapses regardless of $\epsilon$. \emph{Solid lines:} fits of $\Delta r = a\epsilon^n + c$ to all data points, with $c$ absorbing the finite integrator resolution (cf.~text); the fits give $n=0.999$ and $n=1.001$ for periapses and apoapses, respectively.}
\label{fig:scaling}
\end{figure}

The numerical diagnostics of Sec.~\ref{sec:chaos} are illustrative rather than probative.
They exhibit chaos, but chaos alone does not establish that integrability is
lost at \emph{first} order in the tidal coupling $\epsilon$. Indeed, our
equations of motion are exact at $O(\epsilon)$, and the truncated flow is not
integrable at $O(\epsilon^2)$ regardless: generic second-order terms destroy
invariant tori whether or not a Carter-like constant survives at first order.
The numerics could therefore, in principle, be displaying an $O(\epsilon^2)$
effect present independently of the result of Sec.~\ref{sec:integrability}. We
tested this explicitly, using the turning-point structure of the orbits themselves.\footnote{We explain our method for periapses only, but the exact same analysis holds for apoapses with $r_p \rightarrow r_a$ in the exposition. Our results in Fig.~\ref{fig:scaling} show both branches.} 

Consider a single regular tidally-perturbed orbit and follow the total range of all its periapses,
\begin{equation} \label{defDeltarp}
    \Delta r_{p} := \max_{\dot{r}\,=\,0}(r_p) - \min_{\dot{r}\,=\,0}(r_p)\,.
\end{equation}
For a geodesic, the radial turning points are constant, so $\Delta r_p=r_p^{\small \rm geo}-r_p^{\small \rm geo}=0$ by construction, cf.\ Fig.~\ref{fig:turningpoints}. For a tidally-perturbed orbit, the successive turning-point radii differ (cf.\ the right panel of Fig.~\ref{fig:twopoincare}), and $\Delta r\neq0$. The question is whether the way $\Delta r$ scales with $\epsilon$ matches what the non-existence of a Carter constant predicts. It does, and the argument rests on the turning-point structure alone.

Recall first why, for geodesics, the turning points are constant. The two constants of motion $H_0$ and $\Qgeo$ render the Hamilton--Jacobi equation separable~\cite{Schm.02}: the reduced radial motion decouples from the polar one, and the momenta can be written as $p_r^2=R(r;H_0,\Qgeo)$ and $p_\theta^2=\Theta(\theta;H_0,\Qgeo)$, each depending on a single coordinate. Since $\dot r\propto p_r$ (cf.\ footnote~\ref{fn:rdot}), the radial turning points solve $R(r;H_0,\Qgeo)=0$, an equation in $r$ alone: its root(s) depend only on the conserved $(H_0,\Qgeo)$ and are therefore constant along the orbit.

Turning now to the tidally-perturbed dynamics, suppose first that a deformed Carter constant $Q=\Qgeo+\epsilon Q_1$ exists and is in involution with $H=H_0+\epsilon H_1$. Then $(H,Q)$ are two independent constants of motion on the reduced phase space, and by the Liouville--Arnold theorem the motion is again confined to invariant tori, exactly as in the geodesic case but with deformed constants. The inversion applies again,
\begin{equation} \label{eq:radialred}
    p_r^2 = R(r;H,Q) + O(\epsilon^2),
\end{equation}
the entire first-order correction being reabsorbed into the deformed constants $(H,Q)$; any residual $\theta$-dependence is relegated to $O(\epsilon^2)$, an order at which our equations of motion are not controlled anyway. The turning points, fixed by $R(r_p;H,Q)=0$, are thus again constant to first order, and \eqref{defDeltarp} readily becomes
\begin{equation} \label{Delcase1}
    \Delta r_p = O(\epsilon^2).
\end{equation}

Suppose instead that no deformed Carter constant exists. Then $H$ is the only constant of motion, and it can eliminate only one of the two momenta. Eliminating $p_\theta$ and imposing the turning-point condition $p_r\propto\dot{r}=0$ leaves a single relation between the remaining coordinates, and periapses $r_p$ now solve
\begin{equation} \label{eq:rpshift}
    R(r_p;H) + \epsilon\,R_1(r_p,\theta) = 0 + O(\epsilon^2),
\end{equation}
in which the polar angle $\theta$ appears explicitly at $O(\epsilon)$: this is the coupling between radial and polar motion, absent at geodesic order, coming from the non-separability entailed by the non-existence of $Q$. Since $\theta$ varies along the orbit, the periapses inherit this dependence: solving \eqref{eq:rpshift} to leading order gives
\begin{equation} \label{expressionrp}
    r_p(\theta) = r_p^{\small \rm geo} - \epsilon\,\frac{R_1(r_p^{\small \rm geo},\theta)}{R'(r_p^{\small \rm geo})} + O(\epsilon^2),
\end{equation}
and its value differs from one turning point to the next depending no the value $\theta$ reached there. Since $r_p^{\small \rm geo}$ in \eqref{expressionrp} is constant, the periapses \eqref{defDeltarp} is therefore
\begin{equation} \label{Delcase2}
    \Delta r_p = O(\epsilon).
\end{equation}

The exponent in equations \eqref{Delcase1} and \eqref{Delcase2} thus discriminates between the two scenarios: $\Delta r\propto\epsilon^2$ if a deformed Carter constant survives, $\Delta r\propto\epsilon$ if it does not. This is a numerical measurement that allows us to verify if our figures display $O(\epsilon)$-effects due to non-integrability, or $O(\epsilon^2)$-effects due to (uncontrolled) sub-leading physics and numerical artifacts.

Figure~\ref{fig:scaling} shows $\Delta r_p$ and $\Delta r_a$ for a representative regular orbit, at twenty geometrically-spaced values $\epsilon \in [10^{-3},10^{-1}]$. One subtlety must be dealt with. What is measured is $\Delta r_{\rm meas} = \Delta r + \Delta r_0$, where $\Delta r_0$ is the finite resolution of the integrator: a pure geodesic ($\epsilon = 0$) returns $\Delta r_0 \simeq 5 \times 10^{-7}$. Since $\Delta r \sim 10^{-6}$ already at $\epsilon \simeq 10^{-3}$, this floor is not negligible there, and it is not a power of $\epsilon$. Consequently, fitting $\Delta r=a\epsilon^n$ to the full $\epsilon$-range would bias $n$ low,\footnote{Although such a fit returns $n \simeq 0.91$: so even this biased estimate is conclusive, as it favours $n \simeq 1$ over $n \simeq 2$ unambiguously. The refined fit \eqref{eq:drfit} serves to \emph{quantify} the exponent, rather than to establish it.} while fitting only above the floor would discard data on an \emph{a posteriori} criterion. To avoid all this, we therefore fit
\begin{equation} \label{eq:drfit}
    \Delta r_{\rm meas}(\epsilon) = a\,\epsilon^{\,n} + c,
\end{equation}
retaining all data points and letting $c$ describe the aforementioned floor $\Delta r_0$. For the periapsis branch this gives $n = 0.999 \pm 0.005$, $a = (6.98\pm0.01)\times10^{-4}$ and $c = (3.0\pm0.8)\times10^{-7}$, the latter being consistent with the independently measured $\Delta r_0$. The apoapsis branch returns $n = 1.001 \pm 0.004$. The exponent is unity to within half a standard error, and $n=2$ is clearly excluded, cf. the dashed grey line in Fig.~\ref{fig:scaling} showing the corresponding quadratic scaling $\Delta r\propto\epsilon^2$.

The scaling is thus $\Delta r \propto \epsilon$, and the chaos of Sec.~\ref{sec:chaos} reflects a first-order loss of integrability, in quantitative agreement with the analytical result. To summarize the logic: Sec.~\ref{sec:integrability} proves the non-integrability, Sec.~\ref{sec:chaos} illustrates its dynamical consequences, and the present analysis confirms that these consequences are indeed seeded by leading-order tidal effects.

\subsection{Prospects} \label{ssec:prospects}

Several directions follow naturally from this work, and most are already within reach of the tools assembled here.

The first one is astrophysical. We have established that integrability is lost, but not how large its imprint on actual observables is. For an extreme-mass-ratio inspiral, the dimensionless tidal coupling entering the dynamics is minuscule, and KAM theory then guarantees that most invariant tori survive, only slightly deformed, with genuine chaos confined to exponentially thin layers around resonances and to the immediate neighbourhood of the separatrix. The practical imprint on the waveform is therefore unlikely to be problematic, although it would still be worthwhile to make quantitative statements on this, since EMRIs will spend a lot of time (up to $10^5$ radial cycles in the LISA bandwidth. Chaotic effects could also be more plausibly felt at \emph{resonance crossings}, where the fundamental frequencies evolve non-smoothly, and at the loss of a smoothly-defined plunge time, cf.\ Sec.~\ref{sec:chaos}. Whether these effects accumulate into a measurable dephasing over an inspiral, for realistic couplings, is a quantitative question that we have not addressed, and that constitutes the natural next step. It is made tractable by the closed-form Hamiltonian \eqref{Htidalclosed}, in which the perturbation is encoded in three functions of the Weyl scalar alone, so that standard canonical perturbation theory applies away from resonances. A sensible starting point is the simplest non-trivial case: a Schwarzschild background with a quadratic-in-spin quadrupole, which, like the setup studied here, reduces to a two-degree-of-freedom system and is thus amenable to the same phase-space diagnostics we have presented.

The methods themselves are more general than the Kerr-specific case treated here, in two respects. First, the Hamiltonian formulation of Sec.~\ref{sec:Ham} makes no assumption on the background: it holds for a tidally-deformed body in \emph{any} spacetime, and provides a starting point for studies beyond Kerr. Second, the non-existence proof rests on the closed-form tidal invariants \eqref{finalEB}, which we derived using the covariant bivector formalism described in \cite{Ha.20,ComDruVin.23,Ra.Iso.Dru.IntegO2.26}, valid for any spacetime admitting a Killing--Yano tensor. The same machinery therefore applies to several extensions: a spinning \emph{and} tidally-deformed body, in which both (spin and tidal) quadrupoles are considered; dynamical (frequency-dependent) tides, known to matter for neutron-star binaries \cite{StHiDiFo.21}, whose inclusion promotes $(\cE,\cB)$ to dynamical variables and enlarges the phase space; and higher, e.g.\ octupolar, couplings, for which the only new input is the analogue of the closed-form expressions \eqref{finalEB}. Recent works have started looking at the octupole and hexadecapole dynamics in the same Dixon-Harte formalism that we used here \cite{HarteRamond.26,Amancio:2026mnj}. Lastly, we expect our non-existence result itself to extend to the whole family of Einstein spacetimes with a Killing--Yano tensor, studied in \cite{Ra.Iso.Dru.IntegO2.26}.

Finally, on the mathematical side, it should be possible to lift our polynomial-in-momenta Ansatz, although it is quite exhaustive, of our analytic proof. The numerics of Sec.~\ref{sec:chaos} strongly suggest that no smooth integral exists at all, but a global analytic statement, via Melnikov-type or differential-Galois (Morales--Ramis) methods, remains to be established.

\begin{acknowledgments}
I thank S.~Isoyama, A.~Druart, J.~Mathews, M.~Shahzadi, A.~Seenivasan and A.~Le Tiec for discussions. I thank S.~Fauve and J.~le Bourlot for initiating me to dynamical systems and chaos many years ago, and S.~Strogatz for keeping my interest alive since then. I thank Y.~Lemière, F.~Mauger and H.~Alexandre for their remarks and Louis Bernard for his encouragement. This work made use of the \texttt{KerrGeodesics} package of the Black Hole Perturbation Toolkit \cite{BHPToolkit}. Plots and computations were produced with Wolfram \textsc{Mathematica}.
\end{acknowledgments}

\bibliography{ListeRef.bib,ListeRef_Sis.bib,chaos.bib}

\end{document}